\documentclass[%
 aip,
 pof,%
 amsmath,amssymb,
]{revtex4-1}

\pdfoutput=1
\usepackage{amsmath}
\usepackage{graphicx}
\usepackage{subfig}
\usepackage{caption}
\usepackage{dcolumn}
\usepackage{bm}
\usepackage{color}
\usepackage[bottom]{footmisc}

\newcommand{\beqa}{\begin{eqnarray}}
\newcommand{\eeqa}{\end{eqnarray}}
\newcommand{\beq}{\begin{equation}}
\newcommand{\eeq}{\end{equation}}
\newcommand{\be}{\begin{equation*}}
\newcommand{\ee}{\end{equation*}}
\newcommand{\bi}{\begin{itemize}}
\newcommand{\ei}{\end{itemize}}
\newcommand{\ben}{\begin{enumerate}}
\newcommand{\een}{\end{enumerate}}

\newcommand{\dif}{\mathrm{d}}

\begin{document}


\title{A simple stochastic quadrant model for the transport and
  deposition of particles in turbulent boundary layers} 


\author{C.~Jin}
\author{~I.~Potts}%
\author{M. W.~Reeks}
\email{mike.reeks@ncl.ac.uk}
\affiliation{School of Mechanical \& Systems Engineering, Newcastle University, Stephenson Building,
Claremont Road, Newcastle upon Tyne, NE1 7RU, UK}

\date{\today}

\begin{abstract}
We present a simple stochastic quadrant model for calculating the
transport and deposition of heavy particles in a fully developed turbulent
boundary layer based on the statistics of wall-normal fluid velocity
fluctuations obtained from a fully developed channel flow. Individual
particles are tracked through the boundary layer via their interactions
with a succession of random eddies found in each of the quadrants
of the fluid Reynolds shear stress domain in a homogeneous Markov
chain process. In this way we are able to account directly for the
influence of ejection and sweeping events as others have done but
without resorting to the use of adjustable parameters. Deposition
rate predictions for a wide range of heavy particles predicted by
the model compare well with benchmark experimental measurements. In
addition deposition rates are compared with those obtained from continuous
random walk (CRW) models and Langevin equation based ejection and
sweep models which noticeably give significantly lower deposition
rates. Various statistics related to the particle near wall behavior
are also presented. Finally we consider the model limitations in using
the model to calculate deposition in more complex flows where the
near wall turbulence may be significantly different.

\end{abstract}

\maketitle

\section{Introduction}
In this paper, we propose a simple stochastic quadrant model of coherent
structures for heavy particle deposition in a turbulent boundary layer
inspired by the quadrant analysis of \citeauthor{Willmarth1972} \cite{Willmarth1972}
which captures the influence of sweeps and ejections on the deposition
of particles. It is
another way of modeling deposition of heavy particles within fully
developed turbulent boundary layers that adds insight and suggests
new ways for improving the deposition prediction of heavy particles
encountered in a wide range of industrial and environmental applications\cite{friedlander2000smoke}.

Our objective is to show how the influence of ejection and
sweeping events in a turbulent boundary layer on particle
deposition can be taken account of in a simple and more
transparent way than in other models with no adjustable
constants or parameters and at the same time preserving the
statistics of the near wall turbulence.  These are features
of course that make the model highly suited for
implementation in Reynolds-Averaged Navier-Stokes (RANS) CFD
codes like FLUENT or Code\_Saturne for the
prediction of the deposition of heavy particles encountered
in a wide range of industrial and environmental
applications. However, it is the fundamental way this model
deals with the sweeping and ejection events that is the
focus of the study we report here.

The modelling and simulation of the transport and deposition of
particles in a turbulent boundary layer have been and
continue to be much studied topics. The
first attempts of \citeauthor{Friedlander1957}\cite{Friedlander1957}
and \citeauthor{Davies1966}\cite{Davies1966} were based on a gradient
diffusion/free-flight theory where the concept of a particle stop
distance was proposed. However the initial particle free-flight velocity
had to be artificially adjusted from its value based on the local
fluid rms velocity to get good agreement with the experimental data.
\citeauthor{hutchinson1971deposition}\cite{hutchinson1971deposition}
and \citeauthor{Kallio1989}\cite{Kallio1989} employed a Monte-Carlo
based Lagrangian particle tracking method for calculating particle
deposition. In the work of \citeauthor{Kallio1989}\cite{Kallio1989}
the turbulent boundary layer was described as a randomized eddy field
with corresponding velocity and time scales as functions of the particle
distance away from the wall. \citeauthor{swailes1994particle}\cite{swailes1994particle}
proposed to use the kinetic equation developed by \citeauthor{reeks1991kinetic}\cite{reeks1991kinetic}
as a model to study the deposition of ``high inertia'' particles
in a turbulent duct flow. \citeauthor{Young1997}\cite{Young1997}
developed a simple approach based on an advection diffusion equation
(ADE) to address the particle deposition in turbulent pipe flows,
which represents a considerable advance in physical understanding
over previous free-flight theories. \citeauthor{guha1997unified}\cite{guha1997unified}
developed a unified Eulerian theory, which is based on a Reynolds
averaging of the particle continuity and momentum
conservation equations for studying turbulent deposition onto smooth and rough surfaces.
\citeauthor{zaichik2010diffusion}\cite{zaichik2010diffusion} developed
a simplified Eulerian model called the diffusion-inertia model (DIM),
which is based on a kinetic equation for the probability density function
(PDF) of particle velocity and position, to investigate the dispersion
and deposition of low-inertia particles in turbulent flows. Furthermore,
the DIM was incorporated into the nuclear/industrial computational
fluid dynamics (CFD) Code\_Saturne for calculating the deposition of
aerosols (see~\cite{nerisson2011improved}).

 More recently \citeauthor{vanDijk20124904}\cite{vanDijk20124904}
  produced an extremely accurate numerical solution of a PDF
  equation which replicated exactly a random walk model
  simulation of particles in a turbulent boundary layer
  (similar to the statistical model used here but without
  explicitly involving the influence of the ejection and
  sweeps). This is an important step forward because
  observed features such as the concentration
  profiles, are not subject to statistical error or
  numerical inaccuracies and therefore are real effects as
  far as model predictions are concerned.  So we can
  learn a great deal about the mechanisms of non-local
  inertial transport, albeit for a relatively simple random
  walk model, particularly the process of trapping of
  particles in boundary layers and the occurrence of a
  singularity in the particle concentration very close to
  the wall, which makes no contribution to the deposition.

Thanks to significant progress achieved in CFD, and in particular
in the development of sophisticated turbulence models and numerical
methods for unstructured grids for complex geometry, the CFD approach
has been used to study the deposition of heavy particles in both simple
and complex flows and geometries. This is usually carried out in an
Eulerian-Lagrangian framework where individual particles are tracked
through a random Eulerian flow field in which the mean flow, the timescales
and rms of the velocity fluctuations are based on a solution of a
closed set of Reynolds-Averaged Navier-Stokes (RANS) equations for
the underlying carrier-phase flow field. To obtain statistically significant results
it is necessary to carry out the calculation for a huge
number of particles, each
particle associated with a particular realization of the random flow
field. This facility has been embedded into most CFD codes, although
the stochastic nature of both the turbulence of the underlying flow
and the dispersed particulate flow makes the problem of turbulent
dispersed particulate flows more complex than its single-phase counterpart.

Yet prediction of turbulent particle
depositions based on the general RANS modelling framework
still has its shortcomings as demonstrated by numerous
researchers
\cite{greenfield1998numerical,Matida2000,Kroer2000,Tian2007,Dehbi2008,
  horn2008comprehensive, Dehbi2008a, Guingo2008a,
  Chibbaro2008, mehel2010influence} when comparing predictions with the benchmark experimental
measurements of \citeauthor{Liu1974}\cite{Liu1974}. A
particular inadequacy is the isotropic assumption used in
the standard $k-\epsilon$ turbulence model of a general RANS
modelling framework to calculate fluctuating fluid
velocities $u_{i}^{\prime}=\sqrt{2k/3}$.  Associated with
this is the structure and timescale of the near wall
turbulence that is a critically controlling factor for the
deposition of particles. To address this inadequacy,
\citeauthor{Guingo2008a}\cite{Guingo2008a} developed a sophisticated one-dimensional continuous random walk
(CRW) boundary layer model to model the fluid
fluctuating velocity and the interaction of particles with
the near wall coherent structures (e.g. sweeps and
ejections) explicitly. A similar methodology to account for
the interaction of particles with sweeps and ejections has
been employed by
\citeauthor{Chibbaro2008}\cite{Chibbaro2008}, who obtained
satisfactory predictions of deposition rates with the
standard $k-\epsilon$ model in a general CFD modelling
framework. Both \citeauthor{Guingo2008a}\cite{Guingo2008a}
and \citeauthor{Chibbaro2008}\cite{Chibbaro2008}
demonstrated the significant role played by the near wall
coherent structures on the transport and deposition of heavy
particles within turbulent boundary layers.

Since \citeauthor{Kline1967}\cite{Kline1967} first reported the
presence of well-organized spatially and temporally dependent motions
in the near wall region (referred to as bursting) of a turbulent flow,
the role played by coherent structures of near wall on the transport
and deposition of inertia particle has been the focus of attention
of a number of researchers.
\citeauthor{owen1969pneumatic}\cite{owen1969pneumatic} was the first
to suggest that the transport of fine solid particles from a turbulent
gas stream to an adjoining surface may arise from sporadic violent
eruptions from the viscous sublayer. \citeauthor{cleaver1975sub}\cite{cleaver1975sub}
proposed a sub-layer model, which takes into account the role of the
up-sweeps and down-sweeps of fluid observed in the near wall region
of turbulent flows, in order to obtain a better understanding of the
mechanics of the particle deposition process. The model predictions
were in satisfactory agreement with experimental measurements on deposition
rates. The sub-layer model of \citeauthor{cleaver1975sub}\cite{cleaver1975sub}
was used by \citeauthor{fichman1988model}\cite{fichman1988model}
and \citeauthor{fan1993sublayer}\cite{fan1993sublayer} for calculating
particle deposition. \citeauthor{Wei1991}\cite{Wei1991} carried
out a quadrant analysis of laser Dropper anemometer (LDA) measurements
of near wall fluid velocity in order to acquire a preliminary understanding
of suspended sediment transport. \citeauthor{kaftori1995particle_a}\cite{kaftori1995particle_a}
and \citeauthor{kaftori1995particle_b}\cite{kaftori1995particle_b}
demonstrated the importance of coherent wall structures on particle
motion in a turbulent boundary layer and on subsequent entrainment
and deposition processes via a series of systematic experiments. \citeauthor{Marchioli2002}
\cite{Marchioli2002} further examined the mechanisms for particle
transfer and segregation in turbulent boundary layers
through Direct Numerical Simulation (DNS)
calculations of channel flow. They revealed that downward sweeps,
referred to as quadrant IV events, cause particles to transfer to
the near wall region where particles preferentially accumulate in the
low-speed streaks, whilst ejections, referred to as quadrant II events
bring about the migration of particles to the region of outer flow.
\citeauthor{soldati2009physics}\cite{soldati2009physics} provided
a systematic review and physical insight into the physics and modelling
of deposition and entrainment of particles from turbulent
flows which has been a catalyst for better models of particle deposition.

The work of \citeauthor{Wei1991}\cite{Wei1991} has been particularly
important in developing and implementing the methodology for calculating
the particle transport in a turbulent boundary layer proposed in this
paper. They performed the quadrant analysis of \citeauthor{Willmarth1972}\cite{Willmarth1972}
to examine the high-resolution, two-component laser-Doppler anemometer
measurements of the wall normal fluid velocity fluctuations in a fully
developed water channel flow. They found that there is a net upward
momentum flux in the range of $y^{+}>30$ that may be associated with
the bursting process occurring in quadrant II, whilst there is
a net downward momentum flux in the range of $10\le y^{+}\le30$ that
may be associated with the sweeps process occurring in quadrant IV.
The net momentum flux results from the positively skewed distribution
of the fluctuating wall-normal velocity. Inspired by this approach,
the present work proposes another way to model near wall coherent
structures and their interaction with particles under a positively
skewed distribution of  wall-normal fluctuating
velocities. 

It is worth noting that stochastic models of the type considered here
are used to predict particle deposition in complex flows using commercial
CFD codes like FLUENT or the open source Code\_Saturne where they are
used in conjunction with a RANS calculation for the underlying carrier
flow. The near wall flow is based on the application of suitable wall
functions which assume a log-law region and the scaling of the mean
flow and turbulence on a local shear stress (friction velocity) which
is part of the solution of the RANS calculation. The profiles of the
mean flow and the turbulence are thus appropriate for a fully developed
turbulent boundary layer. In incorporating our model here into a RANS
calculation of the carrier flow we would be making the same assumption
about the statistics we employ. It is of course perfectly compatible
with the type of wall function that is being employed but the assumption
of similarity to that of a fully developed turbulent boundary is a
major assumption that needs to be tested.

In formulating and examining this approach, this paper is structured
as follows. First, the stochastic quadrant model is formulated and
discussed. We then present the statistics in the four quadrants obtained
using a quadrant analysis of the wall-normal fluid velocity fluctuations
acquired from an LES of a fully developed channel flow. Finally, results
for the deposition rates from an implementation of this stochastic
quadrant model are presented and compared with results from benchmark
experimental measurements, and those obtained from a one-dimensional
Langevin equation-based CRW model and other CRW models. Several statistics
concerning the transported particles in the near wall region are also
shown. 

\section{Modelling methodology}

\subsection{Governing equations of particle motion}
A Lagrangian particle tracking module was developed and coupled with
an unstructured cell-centred finite-volume based Navier-Stokes equation solver 
to calculate trajectories of heavy particles in flow fields. The focus
of this work is on the deposition of non-colliding, rigid, spherical
and heavy particles. 
For the numerical simulations presented
here the ratio of particle density ($920~\mathrm{kg/m^3}$) to fluid density
is $770$, which is the same as the experimental measurements of \citeauthor{Liu1974}\cite{Liu1974}.
Density ratio is particular to the calculations presented, and not a fundamental part of model.
The concentration of particles is dilute enough to assume one-way
coupling. The particle equation of motion discussed by \citeauthor{maxey1983equation}\cite{maxey1983equation}
is simplified in this work by taking into account only the drag force.
We thus can write the particle equation of motion involving the non-linear
form of the drag law with the point particle approximation 
\beq 
\frac{\dif \mathbf{v_p}}{\dif t} =
\frac{1}{\tau_p}C_D\frac{Re_p}{24}(\mathbf{u} -
\mathbf{v}_p), 
\label{eq:ch4bem}
\eeq 
where $\mathbf{v}_{p}$ is the particle velocity and $\mathbf{u}$ the instantaneous fluid
velocity at the particle position for a particle with
response time $\tau_{p}$. Previous research effort on particle dispersion in a turbulent
channel flow (see\cite{marchioli2006particle}) has demonstrated that
the particle Reynolds number, $Re_{p}=|\mathbf{u}-\mathbf{v}_{p}|d_{p}/\nu$
does not necessarily remain small enough to assume Stokes drag. Thus, an empirical relation
for $C_{D}$ from \citeauthor{Morsi1972}\cite{Morsi1972}, which is applicable to a wide
range of particle Reynolds number with sufficiently high accuracy,
is employed, namely 
\beq 
C_D = c_1 + \frac{c_2}{Re_p} +
\frac{c_3}{Re_p^2},
\label{eq:Cdexpression}
\eeq
in which $c_{1},c_{2},c_{3}$ are constants
and provided by \citeauthor{Morsi1972}\cite{Morsi1972}. The above
empirical expression exhibits the correct asymptotic behavior at low
as well as high values of $Re_{p}$\footnote{A state-of-art composite correlation
for drag coefficient and lift coefficient has been
investigated and is the subject of a subsequent paper.}.

The position $\mathbf{x}_{p}$ of particles is obtained from the kinematic
relationship 
\beq
\frac{\dif \mathbf{x}_p}{\dif t} = \mathbf{v}_p.
\label{eq:ch4kinematicrelation}
\eeq

The boundary condition for
the above equation is that the particle is captured by the wall when
its center is less than its radius away from the nearest
wall. It is worth pointing out here that the present
stochastic quadrant model does not take into account the effect of
build-up of deposited particles on the incoming
particles.  The particle
capture is assumed to be perfectly absorbing with no subsequent re-suspension.

From a converged RANS computation of the velocity flow
field, Eq:~(\ref{eq:ch4kinematicrelation}) is integrated in
time using the second-order Adams-Bashforth scheme to obtain
particle trajectories, whilst Eq:~(\ref{eq:ch4bem}) is
integrated with the second-order accurate Gear2 (backward
differentiation formulae) scheme to obtain instantaneous
velocity of particles. Fluid velocities solved are stored
at the centroids of grid cells on the basis of using
cell-centred finite-volume scheme. Since it is only by
chance that a particle coincides with the cell centroid, a
quadratic scheme based on velocity gradient reconstruction
is used to interpolate the fluid velocity to the particle
location. The collective statistical properties of the
particle phase are obtained by following the trajectories of
$10^{5}$ particles.

\subsection{Formulation of the stochastic quadrant model}

The discrete random walk (DRW, also known as Monte-Carlo eddy interaction)
model is the basis of the formulation of the present stochastic model.
The fluid velocity field in the absence of the dispersed particle
phase is determined by a RANS computation with the standard $k-\epsilon$
model (see~\cite{Wilcox1993}). The temporal fluctuations of the
velocity field are described by a sequence of discrete eddies, with
which the suspended particles interact for a randomized eddy lifetime.
In the particle equation of motion (\ref{eq:ch4bem}), the instantaneous fluid velocity
is represented by a Reynolds decomposition of averaged and fluctuating
components, $\mathbf{u}=\mathbf{\overline{U}}+\mathbf{u}^{\prime}.$
The time-averaged fluid velocity $\mathbf{\overline{U}}$ is acquired
from the solution of a RANS calculation for the turbulent flow. Thus
it is crucial to model the fluctuating components to account for the
effect of turbulence on the dispersion of particles. In this respect,
there have been a number of attempts that we have referred to in the
introduction (see also~\cite{Tian2007,Dehbi2008,Dehbi2008a,Guingo2008a,Chibbaro2008,mehel2010influence}).

In this work, our attention is confined to the turbulent deposition
of particles onto perfectly absorbing adjacent surfaces in a fully
developed turbulent boundary layer, in which the flow velocity statistics
are independent of the streamwise coordinate $x$. As stated
by \citeauthor{pope2000turbulent}\cite{pope2000turbulent},
a fully developed channel flow can be considered as statistically
stationary and one-dimensional, with velocity statistics
depending only
on the wall normal direction $y$. In this case, a new approach
is proposed here to model the wall-normal fluctuating velocity component
denoted by $v^{\prime}$ based on quadrant analysis of the coupled
$(u^{\prime},v^{\prime})$ Reynolds shear stress domain. In addition,
particle tracking is performed using the Lagrangian approach.

It is widely considered that the distribution of the wall normal fluctuating
velocity is skewed within fully developed turbulent boundary layers
(see~\cite{Kim1987}). The wall normal fluctuating velocity
component $v^{\prime}$
can be distinguished as positive or negative according to whether the
momentum flux is away from or towards the wall. Thus let $v_{+}^{\prime}$
be a function defined as

\beq
 v_{+}^{\prime} = \left\{%
 \begin{array}{ll}
 v^{\prime} &\textrm{if }  v^{\prime} > 0,\\
 0 &\textrm{if } v^{\prime} \le 0
 \end{array}%
 \right.
\eeq
\\

and $v_{-}^{\prime}$ defined as

\beq
 v_{-}^{\prime} = \left\{%
 \begin{array}{ll}
 v^{\prime} &\textrm{if }  v^{\prime} < 0,\\
 0 &\textrm{if } v^{\prime} \ge 0.
 \end{array}%
 \right.
\eeq
\\
It is possible to define the average value of $v_{+}^{\prime}$ and
$v_{-}^{\prime}$ as: $\left<v_{+}^{\prime}\right>=\frac{1}{T_{+}}\int_{0}^{T}\, v_{+}^{\prime}\, dt$
and $\left<v_{-}^{\prime}\right>=\frac{1}{T_{-}}\int_{0}^{T}\, v_{-}^{\prime}\, dt,$
where $T$ is the interval of observation time containing the fraction
of $v_{+}^{\prime}$ denoted by $T_{+}$ and the fraction of $v_{-}^{\prime}$
denoted by $T_{-}$. We then have $\left<v_{+}^{\prime}\right>+\left<v_{-}^{\prime}\right>=\frac{1}{T_{+}}\int_{0}^{T}\,\left(v_{+}^{\prime}+\frac{T_{+}}{T_{-}}v_{-}^{\prime}\right)\, dt.$
Accordingly, 
 
\beq
\frac{1}{T} \int_0^{T}\,\left(v_{+}^{\prime} + v_{-}^{\prime}  \right)
\, dt = 0.
\eeq

Thus if $T_{+}<T_{-}$, 
\beq \left|\left< v_{+}^{\prime} \right
>\right| > \left|\left<v_{-}^{\prime}\right>\right|,
\label{eq:commean0} \eeq 
if $T_{+}>T_{-}$, 
\beq 
\left|\left<v_{+}^{\prime}\right>\right| < \left|\left<
    v_{-}^{\prime}\right>\right|.\label{eq:commean1}
\eeq

Similarly, average momentum flux per unit area can be defined as:
\beq 
\left < {v_{+}^{\prime}}^2 \right >
 =  \frac{1}{T_{+}}
\int_0^{T}\,\left({v_{+}^{\prime}}\right )^2 \, \dif t
\label{eq:meanmom0}
\eeq
 
and 
\beq 
\left < {v_{-}^{\prime}}^2 \right >  = 
\frac{1}{T_{-}} \int_0^{T}\, \left ({v_{-}^{\prime}} \right)^2 \, \dif t. 
\eeq
According to Eq:~(\ref{eq:commean0}), when  $T_{+} < T_{-}$
we have  
\beq
\left |\left < {v_{+}^{\prime}}^2 \right > \right| > \left |
\left < {v_{-}^{\prime}}^2 \right > \right |, 
\eeq
and according to Eq:~(\ref{eq:commean1}), when  $T_{+} > T_{-}$
\beq
\left | \left < {v_{+}^{\prime}}^2 \right > \right | < \left
  | \left < {v_{-}^{\prime}}^2 \right > \right |. 
\eeq
It is obvious that $\left |\left < {v_{+}^{\prime}}^3 \right > \right| > \left |
\left < {v_{-}^{\prime}}^3 \right > \right | $ when $T_{+} <
T_{-}$; whilst $\left |\left < {v_{+}^{\prime}}^3 \right > \right| < \left |
\left < {v_{-}^{\prime}}^3 \right > \right | $ when $T_{+} >
T_{-}$.  These two cases mean that the wall normal
fluctuating component is derived from  positively 
and negatively skewed distributions, respectively.  Under the
positively skewed distribution, there will be a net upward
momentum flux of fluid; whilst under the
negatively skewed distribution, there will be a net downward
momentum flux of fluid. This imbalance of momentum flux
of fluid particle within fully turbulent boundary layers
can play an important role on the transport and deposition
of heavy particles. The data from \citeauthor{Kim1987}\cite{Kim1987} show that the
wall normal fluctuating component is of positive skewness in
the range of $0 < y^{+} < 10 $ and $y^{+} > 30$ and of negative skewness in
the range of $10 < y^{+} < 30$. 

\subsection{Statistics of $v^{\prime}$ in each of the four quadrants}
 
Following the quadrant analysis approach of \citeauthor{Willmarth1972}
\cite{Willmarth1972} for analysing the structure of the Reynolds
stresses, we classified the wall normal fluctuating velocity and averaged
it in the four quadrants according to the instantaneous flow velocity
in the quadrant domain. In this sense, the instantaneous velocity
of a sufficiently large number of fluid particles at a specified position
may be categorized in terms of the sign of the streamwise and wall
normal velocity fluctuations. For example, when both $u^{\prime}$
and $v^{\prime} > 0$ the instantaneous velocity signal is allocated
to quadrant I ($Q_{\mathrm{I}})$; in the case of $u^{\prime}<0$ and $v^{\prime}>0$,
it is allocated in quadrant II $(Q_{\mathrm{II}})$; when both $u^{\prime}$
and $v^{\prime}<0$ , is allocated to the quadrant III $(Q_{\mathrm{III}})$;
finally, if $u^{\prime}>0$ and $v^{\prime}<0$, it is allocated to
quadrant IV $(Q_{\mathrm{IV}})$. We note that upward momentum fluxes may be
associated primarily with the bursting process associated with events
in $Q_{\mathrm{II}}$, whilst downward momentum fluxes are associated with
sweep events in $Q_{\mathrm{IV}}$. Physically speaking, upward momentum fluxes
associated with $Q_{\mathrm{II}}$ would cause particles to move away from
the wall and downward momentum fluxes associated with $Q_{\mathrm{IV}}$ would
result in the migration of particles toward the wall.

Time averages of $v^{\prime}_i$ and momentum flux $v^{\prime}{}^{2}_i$
can be defined for each of the four quadrants according to Eq:~\ref{eq:commean0}
and ~\ref{eq:meanmom0} as 
\beq
\left < {v'}_{i} \right> = \frac{1}{T_i}\int_0^T
\,{v'}_{i}\, \dif t; \qquad i = \text{I, II, III, IV} 
\eeq
and 
\beq
\left < {v'}_{i}^2 \right>   = \frac{1}{T_i}\int_0^T
\, {v'}_{i}^2\, \dif t; \qquad i = \text{I, II, III, IV,}
\eeq
where $T_i$ denotes time spell spent in the quadrant $i$ by ${v'}_{i}$,  and ${v'}_{i}$ is define as
\beq
 v_{i}^{\prime} = \left\{%
 \begin{array}{ll}
 v^{\prime} &\textrm{if }  v^{\prime} \textrm{ satisfies the
 criterion of quadrant analysis,}\\
 0 &\textrm{if not}.
 \end{array}%
 \right.
\eeq

\begin{figure}[htp]
  \centering
  \includegraphics[trim=0.00in 0.0in 0.0in
  0.0in,clip=true,scale=1.0, angle = 0, width = 1.0\textwidth]{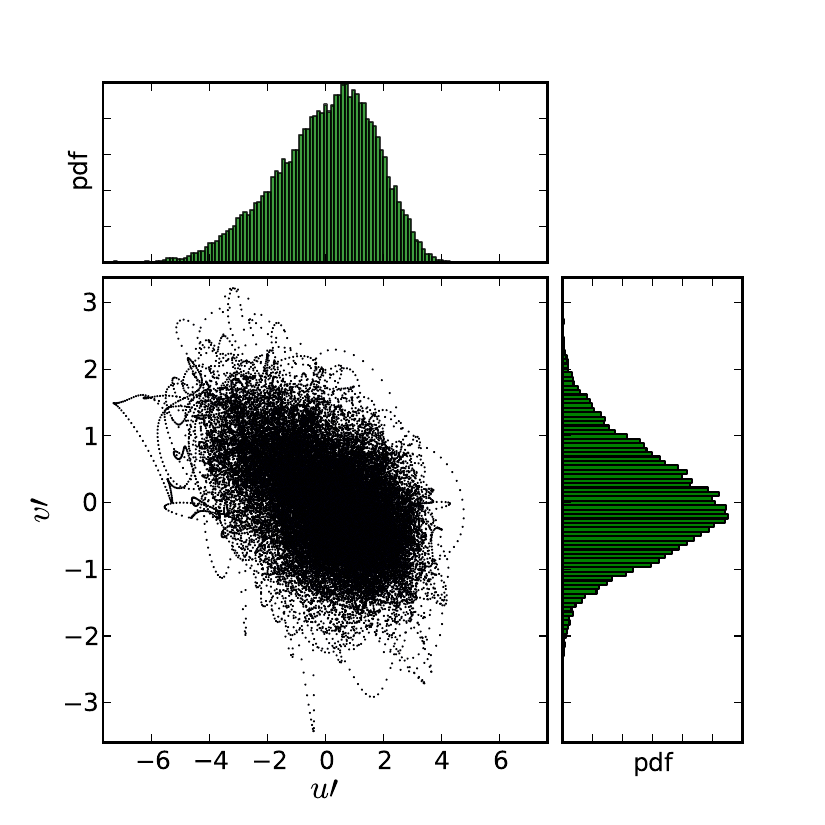} 
  \caption[Scatter plot of $u^{\prime}$ and $v^{\prime}$ at
  $y^{+} = 50 $ resolved by LES, categorised in terms of
  quadrant analysis.]
{Scatter plot of $u^{\prime}$ and $v^{\prime}$ at
  $y^{+} = 50 $ resolved by LES, categorised in terms of quadrant analysis.}
\label{fig:uv_dash_y50}
\end{figure}

A large eddy simulation (LES) of a fully developed channel flow
with $Re_{\tau}=180$ was carried out to obtain the corresponding
statistics of $v_{i}^{\prime}$ up to $y^{+}=100$. The LES was based
on a dynamic Smagorinsky sub-grid scale (SGS) model \cite{germano1991dynamic}
and a generalized fractional-step method\cite{kim1985application}
for the overall time-advancement. A scatter plot of $u^{\prime}$
and $v^{\prime}$ together with the corresponding probability density
function (pdf) (statistically integrated over a non-dimensional time unit) is shown
in FIG.~\ref{fig:uv_dash_y50} according to the quadrant
analysis\cite{Willmarth1972}.
We find that the probability density functions of $u^{\prime}$
and $v^{\prime}$ are both skewed.

\begin{figure}[htp]
  \centering
  \includegraphics[trim=0.00in 0.0in 0.0in
  0.0in,clip=true,scale=1.0, angle = 0, width =
  1.0\textwidth]{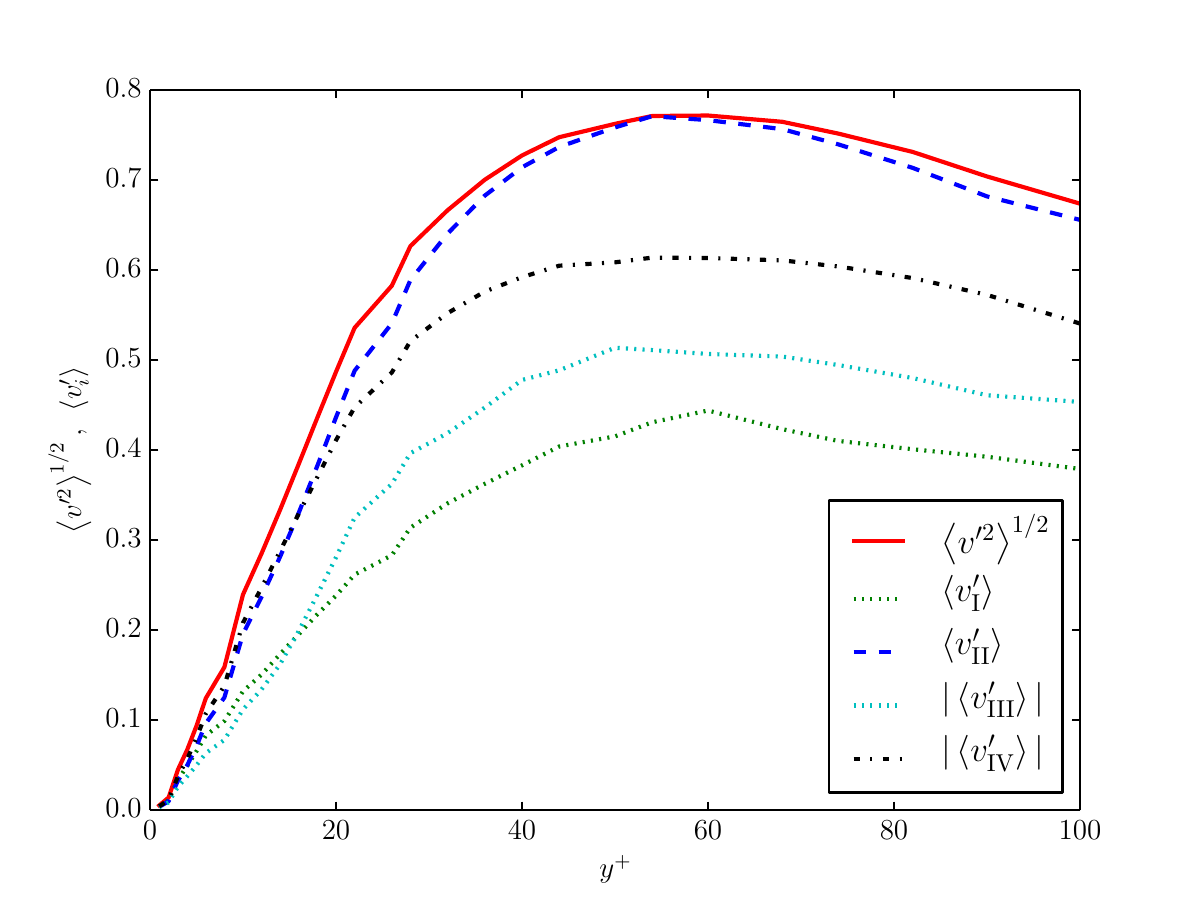} 
  \caption[Profiles of $\left<v^{\prime 2}\right>^{1/2}$ and $\left< {v'}_{i}
    \right>$ as a function of $y^{+}$ at $Re_{\tau} = 180$
    in each of the four quadrants.]
{Profiles of $\left<v^{\prime 2}\right>^{1/2}$ and $\left< {v'}_{i}
    \right>$ as a function of $y^{+}$ at $Re_{\tau} = 180$ in each of the four quadrants.}
\label{fig:vprime0}
\end{figure}

\begin{figure}[htp]
  \centering
  \includegraphics[trim=0.00in 0.0in 0.0in
  0.0in,clip=true,scale=1.0, angle = 0, width = 1.0\textwidth]{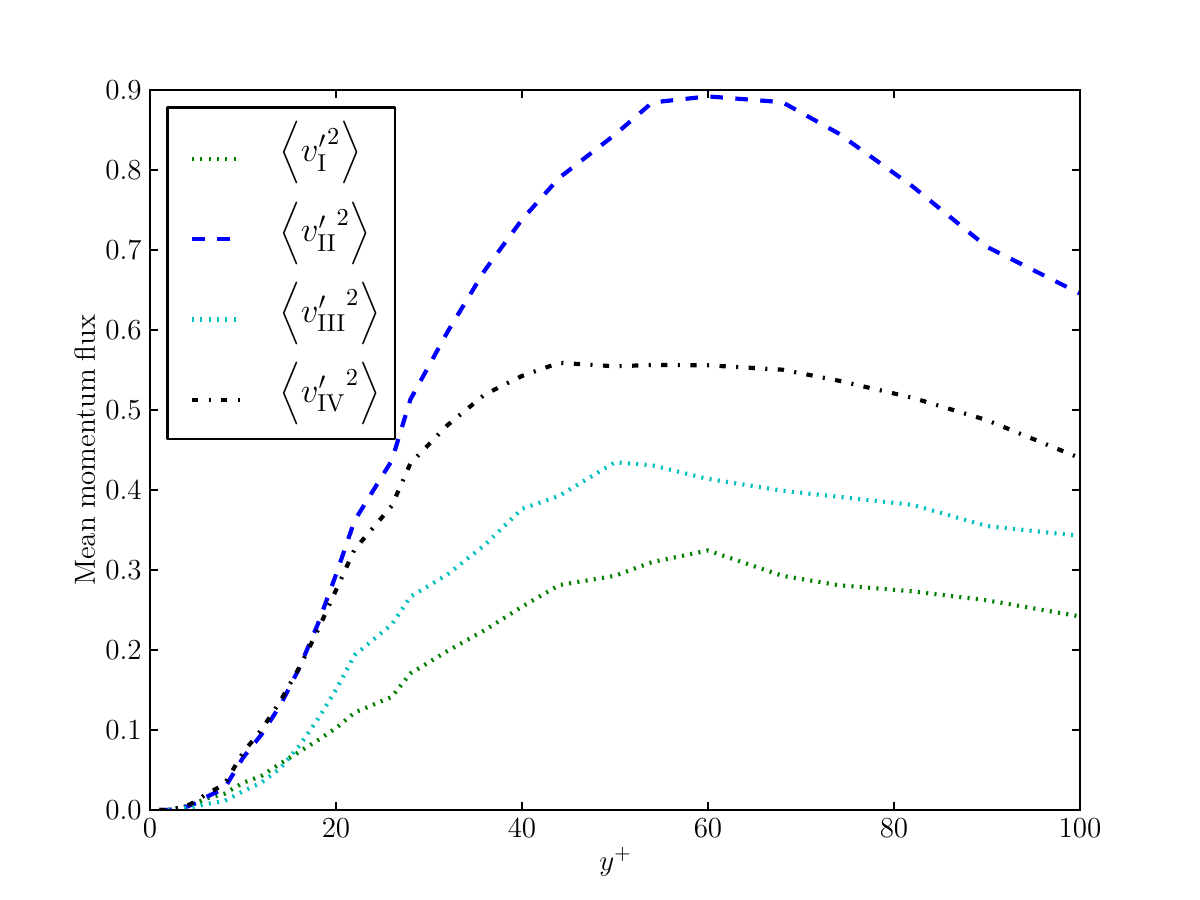} 
  \caption[Profiles of momentum fluxes as a function of
  $y^{+}$ at $Re_{\tau} = 180$ in each of the four quadrants.]
{Profiles of momentum fluxes as a function of $y^{+}$ at
  $Re_{\tau} = 180$ in each of the four quadrants.}
\label{fig:momentumflux0}
\end{figure}

In FIG.~\ref{fig:vprime0}, the quadrant mean $\left<v_{i}^{\prime}\right>$
and wall normal flow velocity rms $\left<v^{\prime 2}\right>^{1/2}$
as a function of $y^{+}$ show that the fluctuating components in
the four quadrants are smaller in magnitude than the wall
normal flow velocity rms $\left<v^{\prime 2}\right>^{1/2}$
across the $y^{+}$ range shown. $\left<v_{i}^{\prime}\right>$
in each of the four quadrants is of different magnitude, indicating
that there is an asymmetry in the wall normal fluctuating components.
Furthermore, the greatest magnitude of $\left<v_{i}^{\prime}\right>$
is found in $Q_{\mathrm{II}}$ across most of the $y^{+}$ range. FIG.~\ref{fig:momentumflux0}
shows that there is a net upward momentum flux resulting from $Q_{\mathrm{II}}$
for the range of $y^{+}>20$. However, this situation is reversed in
the range of $y^{+}<20$. The asymmetry of $\left<v_{i}^{\prime}\right>$
and $\left<v_{i}^{\prime}{}^{2}\right>$ in each of the four quadrants
is a new feature for modelling velocity fluctuations encountered by
heavy particles, which we believe is particularly useful for measuring
particle transport and deposition in the near wall region.

\subsection{Implementation of the   stochastic quadrant model}

The imbalance of $\left<v_{i}^{\prime}\right>$ within each of the
four quadrants will be of differing importance to the transport and
deposition of heavy particles. Events in quadrant II are mainly associated
with violent ejections of low-speed fluid away from the wall; motions
in quadrant IV are primarily associated with an inrush of high-speed
fluid toward the wall, also referred to as the sweeping event. There
are no significant structures associated with quadrant I and III.
The upward momentum flux in quadrant II may be a strongly contributing
factor in the transport of particles away from the wall, reducing
the deposition rates; whilst the inward momentum flux in quadrant
IV may be a strongly contributing factor in the transport of particles
towards the wall, tending increase the deposition rates.

\begin{figure}[htp]
\centering
\includegraphics[trim=0.00in 0.0in 0.0in 0.0in,clip=true,scale=1.0, angle = 0, width = 1.0\textwidth]{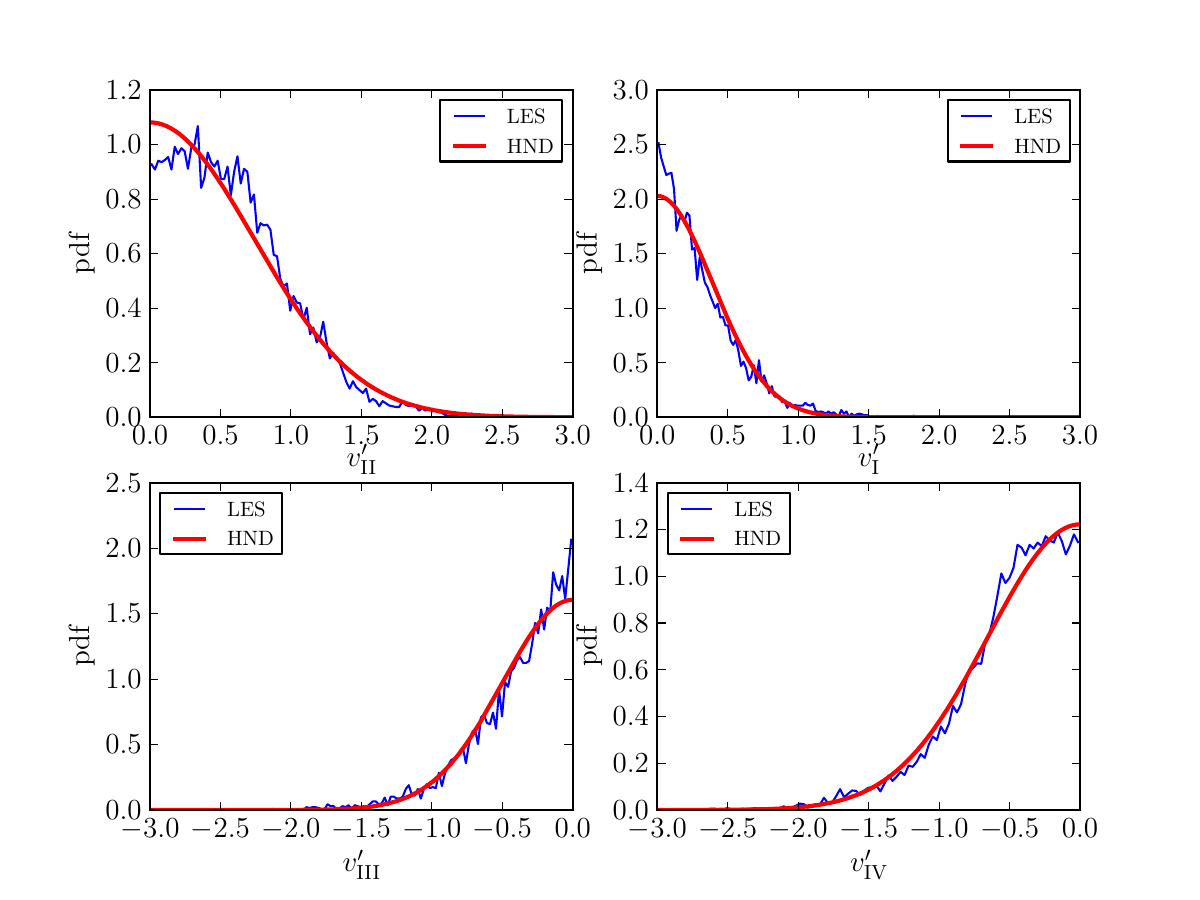} 
\caption[Probability density function (pdf) for $ v_{i}^{\prime} $ obtained by LES and a comparison
with a half normal distribution (HND)]
{Probability density
  function (pdf) for $ v_{i}^{\prime} $
  obtained by LES and a comparison with a half normal
  distribution (HND) }
\label{fig:pdf_hnd}
\end{figure}

The results on $\left<v_{i}^{\prime}\right>$ and $\left<v_{i}^{\prime}{}^{2}\right>$
enables us to specify the statistics of wall-normal velocity fluctuations
encountered by particles in each  eddy along their trajectories.
For example, curve-fitting of the four profiles of $\left<v_{i}^{\prime}\right>$
can be achieved easily. However, comparing the shape of $\left<v_{i}^{\prime}\right>$
with the shape of $\left<v^{\prime}{}^{2}\right>^{1/2}$, a different
scale factor is assumed between $\left<v_{i}^{\prime}\right>$ and
$\left<v^{\prime}{}^{2}\right>^{1/2}$. In FIG.~\ref{fig:pdf_hnd}
the probability density functions for a half normal distribution for
$v_{i}^{\prime}$ in each of the four quadrants at $y^{+}=30$
are shown to be in fair agreement with the original LES results, indicating that a
half normal distribution may be used to describe the distribution
of $v_{i}^{\prime}$. This probability distribution function is given
by 
\beq
 f_{\Omega}(\omega;\,\sigma) = \left\{%
 \begin{array}{ll}
 \frac{\sqrt{2}}{\sigma
  \sqrt{\pi}}\mathrm{exp}\left(-\frac{\omega^2}{2\sigma^2}\right)
&\textrm{if } \omega \ge 0,\\
 0 &\textrm{if } \omega < 0,
 \end{array}%
 \right.
\eeq

where $\sigma$ is set to equal to the value of $\sqrt{\frac{\pi}{2}}\left<v_{i}^{\prime}{}^{2}\right>^{1/2}$
at the corresponding $y^{+}$ location.

\begin{figure}[htp]
\centering
\includegraphics[trim=3.70in 5.2in 3.1in
  4.2in,clip=true,scale=0.95, angle = 0, width = 1.0\textwidth]{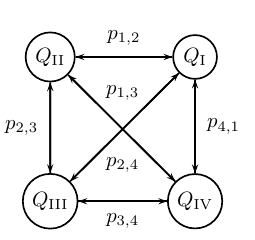} 
\caption[Diagram describing the Markov chain modelling
motions in the four quadrants.]
{Diagram describing the Markov chain modelling
motions in the four quadrants.}
\label{fig:diagram_mc}
\end{figure}

The logical next step is to construct a random process, which models
the eddy motions in the four quadrants. Particles will interact with
a random succession of eddies resulting from different quadrants.
For this, a homogeneous Markov chain was conceived as a model for
the evolution of eddy events in the four quadrants along the particle
trajectories. Particles may interact with an eddy in quadrant I. After
this eddy decays, they would then be able to interact with an eddy
resulting from any of the four quadrants with a certain transition
probability. FIG.~\ref{fig:diagram_mc} describes this process.

As far as the transition probabilities are concerned, let $Q_{i},i=\{\mathrm{I},\mathrm{II},\mathrm{III},\mathrm{IV}\}$
be a discrete time Markov chain on $\{Q_{\mathrm{I}},
Q_{\mathrm{II}}, Q_{\mathrm{III}}, Q_{\mathrm{IV}}\}$
with a transition matrix 

\beq
P = 
\begin{pmatrix}
p_{11} & p_{12} & p_{13} & p_{14}\\
p_{21} & p_{22} & p_{23} & p_{24}\\
p_{31} & p_{32} & p_{33} & p_{34}\\
p_{41} & p_{42} & p_{43} & p_{44}
\end{pmatrix},
\label{eq:dsmc_matrix}
\eeq

where ($p_{ij}:i,j\in\{1,2,3,4\}$) denotes the corresponding
probability distribution of random eddy events in each quadrant. The
transition matrix in Eq:~\ref{eq:dsmc_matrix}, needs to satisfy
the condition ${\displaystyle \sum_{j}p_{ij}=1}$. For eddy events
in the four quadrants, Eq:~\ref{eq:dsmc_matrix} is reduced to
a {}``degenerate'' transition matrix as 
\beq
P = 
\begin{pmatrix} p_{11} & p_{22} & p_{33} & p_{44}
\end{pmatrix}.
\label{eq:dsmc_matrix1}
\eeq  

\begin{figure}[htp]
\centering
\includegraphics[trim=0.00in 0.0in 0.0in
  0.0in,clip=true,scale=1.0, angle = 0, width = 1.0\textwidth]{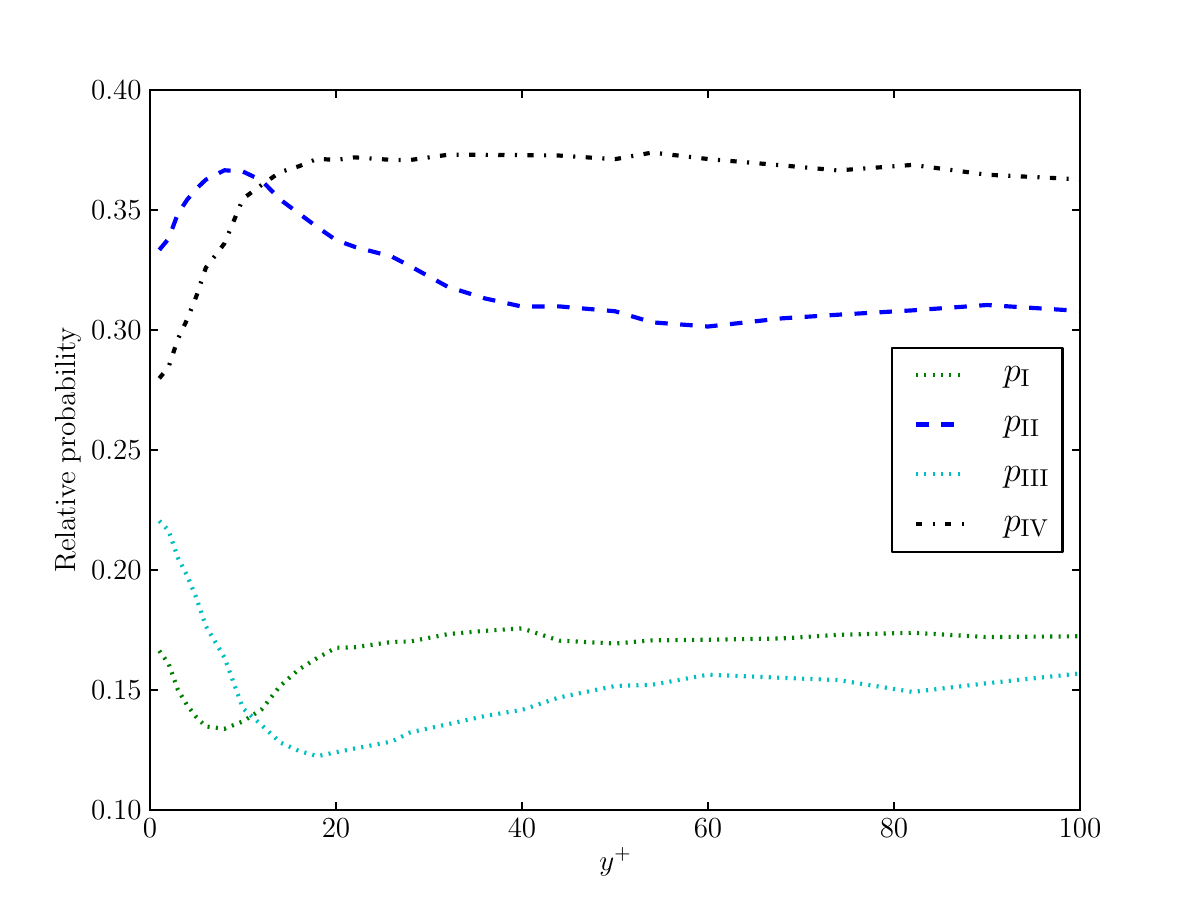} 
\caption[Relative probability of four quadrants as a
function of $y^{+}$.]
{Relative probability of four quadrants as a
function of $y^{+}$.}
\label{fig:relative_probability}
\end{figure}

FIG.~\ref{fig:relative_probability} shows variations of the
relative probability associated with each of the four quadrants as
a function of $y^{+}$, which are computed in terms of the
sign of each individual event within the integrated
non-dimensional time. These probabilities are used as the transition
probabilities denoted in Eq:~(\ref{eq:dsmc_matrix1}).

\begin{figure}[htp]
\centering
\includegraphics[trim=0.00in 0.0in 0.0in
  0.0in,clip=true,scale=1.0, angle = 0, width = 1.0\textwidth]{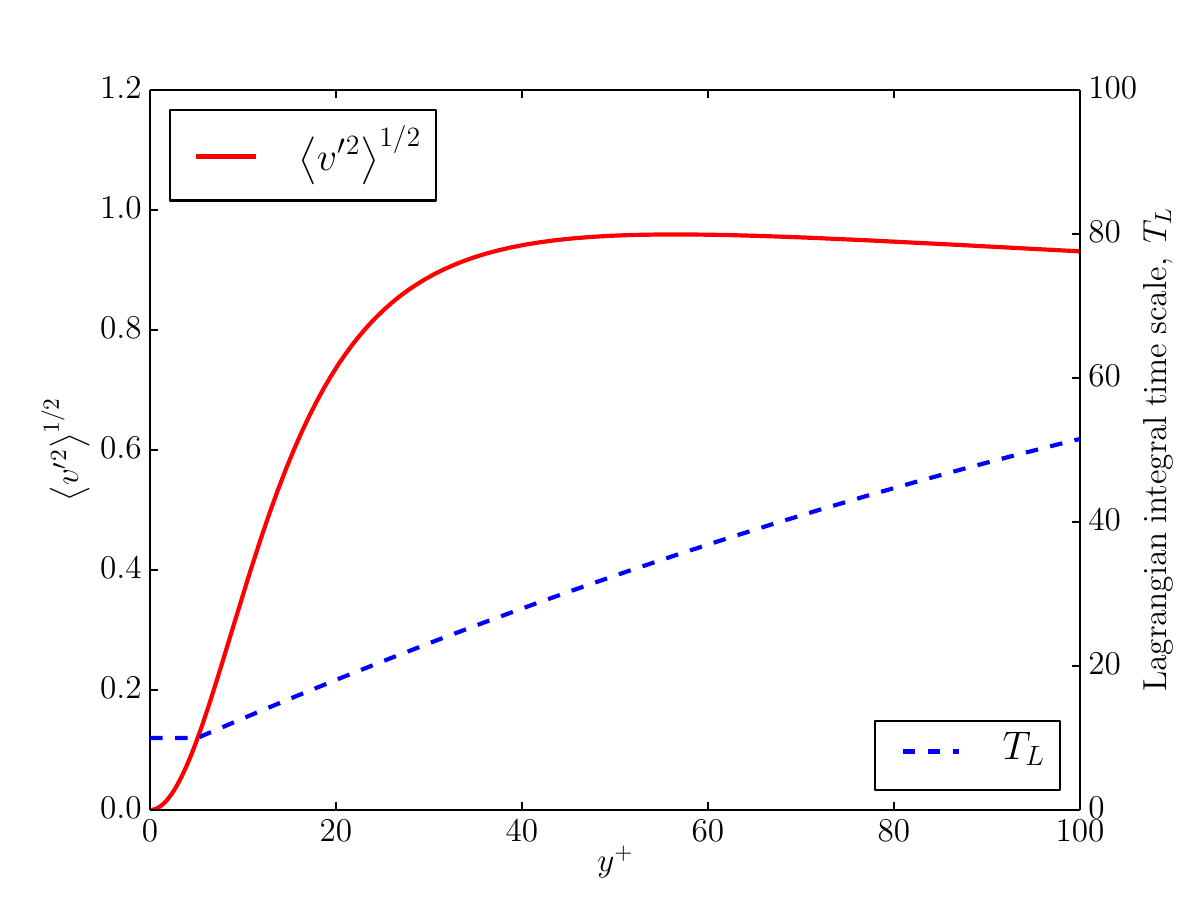} 
\caption[Non-dimensional wall normal fluid velocity and
Lagrangian time integral time scale as a function of
$y^{+}$ within turbulent boundary layers.]
{Non-dimensional wall normal fluid velocity and
Lagrangian time integral time scale as a function of
$y^{+}$ within turbulent boundary layers.}
\label{fig:lagrangian_time_scale}
\end{figure}

The time scale of eddies in each of the four quadrants is difficult
to estimate accurately from the present study, although \citeauthor{tiederman1987timescale}\cite{tiederman1987timescale}
provided several quantitative techniques to measure time scales associated
with bursting events. In the present study, the lifetime of eddies in
the four quadrants are assumed to equal to the Lagrangian time scale
of fluid particles according to their corresponding $y^{+}$ position.
FIG.~\ref{fig:lagrangian_time_scale} shows the Lagrangian time scale
of fluid particles within turbulent boundary layers. This is taken
from the curve-fitting of \citeauthor{Kallio1989}\cite{Kallio1989}.
Furthermore, the random Lagrangian time scale $\xi$ is assumed to obey an exponential
distribution with the following PDF
\beq 
f(\xi,T_L) = \frac{1}{T_L} e^{-\frac{\xi}{T_L}} \qquad \xi >
0, 
\eeq
where $T_{L}$ indicates the integral Lagrangian time scale
in wall units at the particle position. FIG.~\ref{fig:lagrangian_time_scale}
also shows the wall-normal rms profile of fluid velocity. $\left<v_{i}^{\prime}\right>$
in each of the four quadrants is obtained by multiplying $\left<v^{\prime}{}^{2}\right>^{1/2}$
by a scaling factor that is the ratio of the magnitude of the velocity
fluctuation in each of the four quadrants to the magnitude of wall-normal
velocity fluctuation across the boundary layer. In every eddy generated in the four quadrants,
the fluctuating velocity is sampled from a half normal distribution
with a mean $\left<v_{i}^{\prime}\right>$ and a variance corresponding
to the particular particle $y^{+}$ value in the boundary
layer.

The present stochastic model used to
  predict the deposition rates of inertial particles can be
  summarized in terms of the following steps: firstly, a
  RANS simulation is run to establish a steady flow field;
  secondly, velocity fluctuations encountered along the
  particle trajectories are generated within the discrete
  Markov chain Monte-Carlo process according to the relevant
  statistics; thirdly, Eqs:{~\ref{eq:ch4bem} and
    \ref{eq:ch4kinematicrelation}} are integrated to
  determine the particle trajectories; finally, the
  deposition rate can be calculated according to the
  penetration efficiency.

\section{Results and discussion}

\subsection{Continuous  phase}

\begin{figure}[htp]
\centering
\includegraphics[trim=0.00in 0.0in 0.0in
  0.0in,clip=true,scale=1.0, angle = 0, width = 1.0\textwidth]{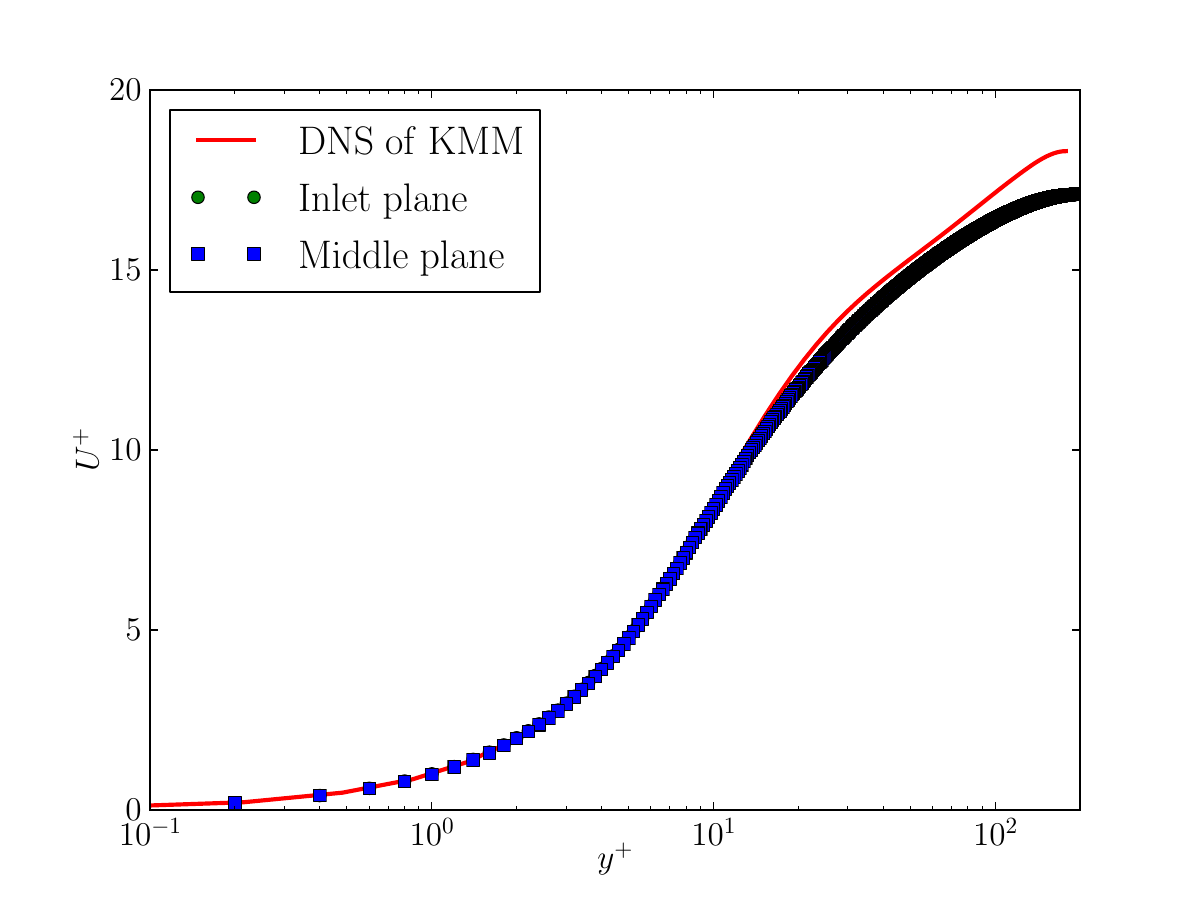} 
\caption[Mean fluid velocity profiles from the inlet and
middle plane.]
{Mean fluid velocity profiles from the inlet and
middle plane.}
\label{fig:uplus_yplus}
\end{figure}

The turbulent boundary layer was resolved using the standard $k-\epsilon$
model with enhanced wall treatment. The $y^{+}$ value
of the first cell adjacent to the wall was set at $y^{+}=1$. Two
points need to be made here. Firstly, there is no discernible discrepancy
between the inlet and middle plane velocity
profiles. Secondly, although the
calculated wall-friction velocity normalized velocity profiles show under-predicted values
when compared with the
DNS data of \citeauthor{Kim1987}\cite{Kim1987}, they show
a very good agreement with the
DNS data of \citeauthor{Kim1987}\cite{Kim1987} when $ y^{+} <
10 $. Given the fact that RANS was employed, the small difference
between the calculated and DNS values shown in
FIG.~\ref{fig:uplus_yplus} for $ y^{+} > 10 $
is reasonable. To avoid a transition region near the inlet
region, an established velocity profile was injected as
velocity boundary condition on the the inlet plane.

\subsection{Dispersed particle phase}
\subsubsection{Particle deposition rates}
  The particle deposition rate in a turbulent boundary layer is usually
quantified through a mass transfer coefficient $K$ defined
as 
\beq
K = \frac{J_{w}}{\overline{c}}, 
\eeq 
where $J_{w}$ represents
the particle flux onto the wall per unit area and time and
$\overline{c}$ is the average particle concentration within the boundary
layer. The computation technique proposed by \citeauthor{Kallio1989}\cite{Kallio1989}
was used to calculate the non-dimensional particle deposition velocity
defined as 
\beq
V_{dep}^{+} = \frac{\overline{U}A}{u_{\tau} P \Delta x}\mathrm{ln}\left(\frac{N_{in}}{N_{out}}\right)
\label{eq:deposition}
\eeq where $\overline{U}$ is the average streamwise fluid
velocity across the fully developed turbulent boundary
layer, $A$ is the boundary layer cross sectional area, $P$
the duct perimeter, $\Delta x$ is the incremental length of
section considered, and $N_{in}$ and $N_{out}$ are the total
number of particles passing through the start and end plane
of each section, respectively. The characteristic wall
friction velocity $u_{\tau}$ was used to obtain the
non-dimensional deposition velocity $V_{dep}^{+}$. On the
basis of Eq:~\ref{eq:deposition}, it can be inferred that
the non-dimensional particle deposition velocity from the
calculated fluid mean velocity profiles shown in
FIG.~\ref{fig:uplus_yplus} will be under-predicted when
compared with the theoretical values. However, given the
small difference shown in FIG.~\ref{fig:uplus_yplus}, the
under-prediction will not significantly affect the
final accuracy of the calculation of the non-dimensional
particle deposition velocity.  In this study, $10^{5}$
particles are introduced uniformly from the inlet plane and
their velocities are initialized  to the value of
the local fluid mean velocity.

\begin{figure}[htp]
\centering
\includegraphics[trim=0.00in 0.0in 0.0in
  0.0in,clip=true,scale=0.75, angle = 0, width = 1.0\textwidth]{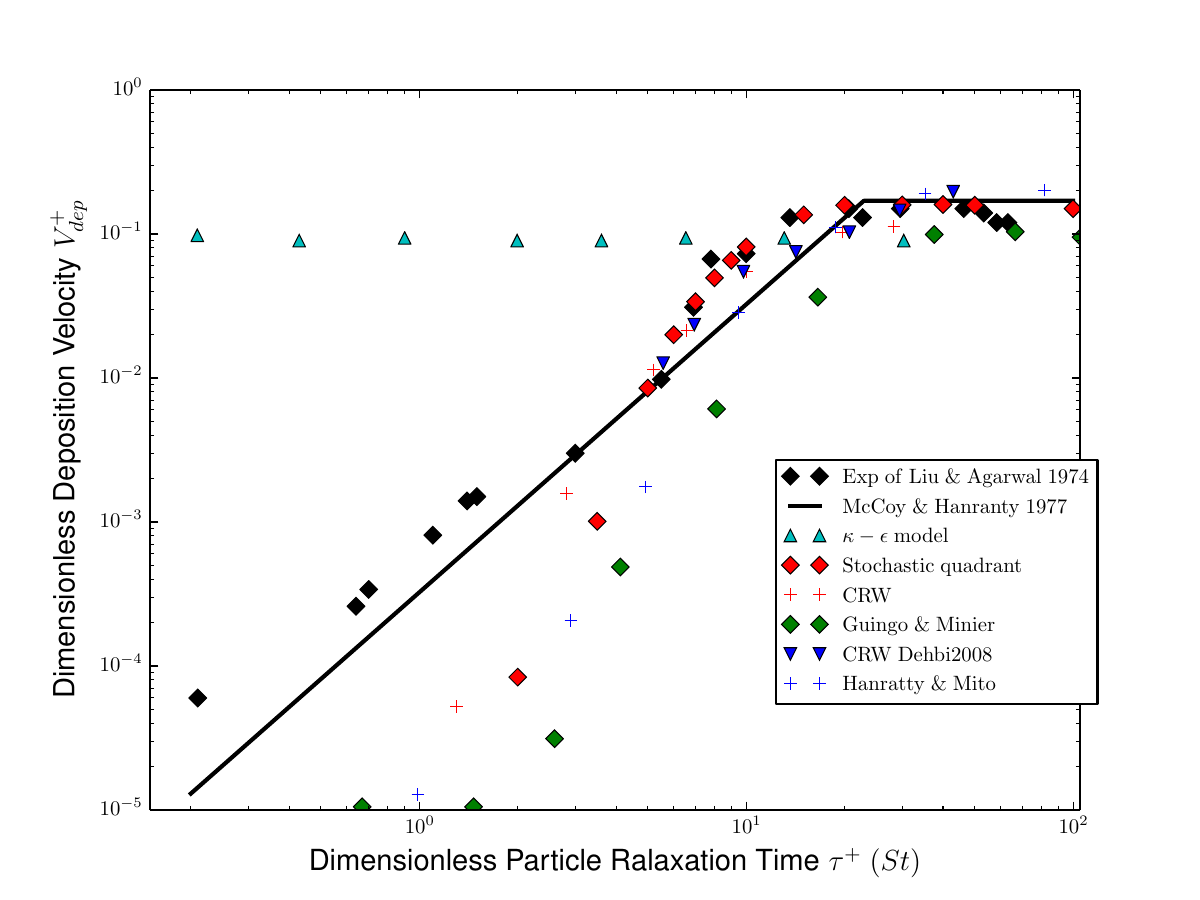} 
\caption[Comparison of dimensionless particle deposition
velocity as a function of dimensionless particle response
time with experimental measurements and different models in
turbulent boundary layers.]
{Comparison of dimensionless particle deposition
velocity as a function of dimensionless particle response
time with experimental measurements and different models in
turbulent boundary layers.}
\label{fig:comparison_vplus0}
\end{figure}

Computed dimensionless particle deposition velocities are compared
with the benchmark experimental measurements (see~\cite{Liu1974}),
the empirical curve-fit of \citeauthor{McCoy1977}\cite{McCoy1977}
and the deposition velocities predicted by the standard $k-\epsilon$ model in FIG.~\ref{fig:comparison_vplus0}.
We have also included for comparison the results from \citeauthor{Guingo2008a}\cite{Guingo2008a}, who developed a sophisticated stochastic model to account for the
geometrical structures in turbulent boundary layers and which like
the current model accounts for the influence of sweeping and ejection
events. It can be observed that very good agreement exists between
the present computed results and experimental data in the range of
$St>5$. For $St<5$, the stochastic quadrant model gives an under-prediction
of the deposition rates. These are also features of the stochastic
model of \citeauthor{Guingo2008a}\cite{Guingo2008a} (G\&M) which predicts significantly 
less deposition than obtained in this study throughout the entire range of $St$. 

In addressing the reasons for the lack of agreement between the predictions
of the present model and that of the G\&M sweep and ejection
model, it is worth considering
some of the critical assumptions that were made by G\&M in the formulation
of their model. Referring to FIGs. 15 and 16 of \citeauthor{Guingo2008a}\cite{Guingo2008a}, we
note that the turbulent boundary layer is divided into outer and inner
zones: in the outer zone a fluid point is transported by a random
succession of coherent structures which model the influence of ejections
and sweeps together with phases in which the flow velocity field is
more random and less persistent and described by a simple Langevin
diffusion process; in the inner zone closest to the wall there are
no coherent structures and the fluid point motion is driven by diffusion.
What is important is that introducing 1D coherent structures into
the outer zone should be consistent with the measured statistics of
the turbulent boundary layer. FIG. 15 of \citeauthor{Guingo2008a}\cite{Guingo2008a} shows the comparison of the
G\&M model predictions of the turbulent kinetic energy $k$ and those
obtained from DNS. There is a noticeable difference in the near wall
region where the G\&M profile of $k$ is significantly steeper. The
authors indicate that the profile of $k$ depends markedly on the
position of the interface between the outer and inner zones (as one
might expect). In their simulation $y_{int}$,  which is the
interface distance away from the wall, varies randomly between
$5$ and $20$ wall units. Better agreement with DNS measurements for
$k$ are obtained if $5<y_{int}<40$ (see FIG. 16 of
\citeauthor{Guingo2008a}\cite{Guingo2008a} ). However in this
case the predicted deposition rates are significantly worse (as one
might expect since the curve is less steep). Despite the fact that
the $k$ profile is noticeably different from that obtained using
DNS, G\&M stick with their initial choice for $y_{int}$ because it
gives better deposition results and consistent with their original
assumption of the location of the sweeping and ejection events. The
steeper profile of $k$ compared with the DNS measurements may be regarded
as an inevitable artifact and penalty one pays for using a simple
model of 1D coherent structures. The model we have presented here
contains the influence of ejection and sweeps reflected in the skewness
of the measured DNS statistics and as such does not introduce spurious
features into the deposition mechanisms. Therefore it is not subject
to fine tuning of model parameters.

In this work, an alternative continuous random walk (CRW) model is
used to repeat the numerical study of particle deposition rates. This
model is based on the wall-normal component of normalized Langevin
equations in boundary layers (see \cite{mito2002use,Dehbi2008a}),
which takes into account the effect of Stokes number along heavy particle
paths (see \cite{Bocksell2006}). The Langevin equation is solved
using a second-order accuracy Milshtein scheme (see \cite{mil1979method}).
The non-dimensional fluctuating fluid velocity solved this way was
converted to a physical velocity and then added into the particle
equation of motion to account for the turbulence effect.

Results for the deposition rates from this CRW model are
also shown in
FIG.~\ref{fig:comparison_vplus0}. There are a few interesting
points to note. Firstly, very similar results are obtained
by us from 
this one-dimensional CRW model compared to those obtained
using CRW by \citeauthor{Dehbi2008a}\cite{Dehbi2008a}.
Secondly, the numerical results from all the models show fair agreement
with experiments for large particles. However, they all give significant
under-predictions on deposition rates for small particles. In contrast
to the present one-dimensional CRW and stochastic quadrant model,
the CRW model employed by \citeauthor{Dehbi2008a}\cite{Dehbi2008a}
was solved in three dimensions with curve-fitting DNS database. This
may further corroborate the view that the wall-normal fluid fluctuations
are a critical control factor on the deposition of heavy particles
from fully developed turbulent boundary layer. Thus, as far as practical
applications are concerned, it is possible and feasible to feed in
only the wall-normal fluid fluctuations for studying particle deposition.
On the other hand, compared to CRW models, the stochastic quadrant
model is capable of yielding equally reliable results for deposition
rates, given its relatively simple nature and the way it
takes account of the influence of sweeps and ejections
in turbulent boundary layers. As a consequence, it is potentially
a very promising model for studying deposition of heavy particles
from turbulent flows.

\begin{figure}[htp]
\centering
\includegraphics[trim=0.00in 0.0in 0.0in
  0.0in,clip=true,scale=0.75, angle = 0, width = 1.0\textwidth]{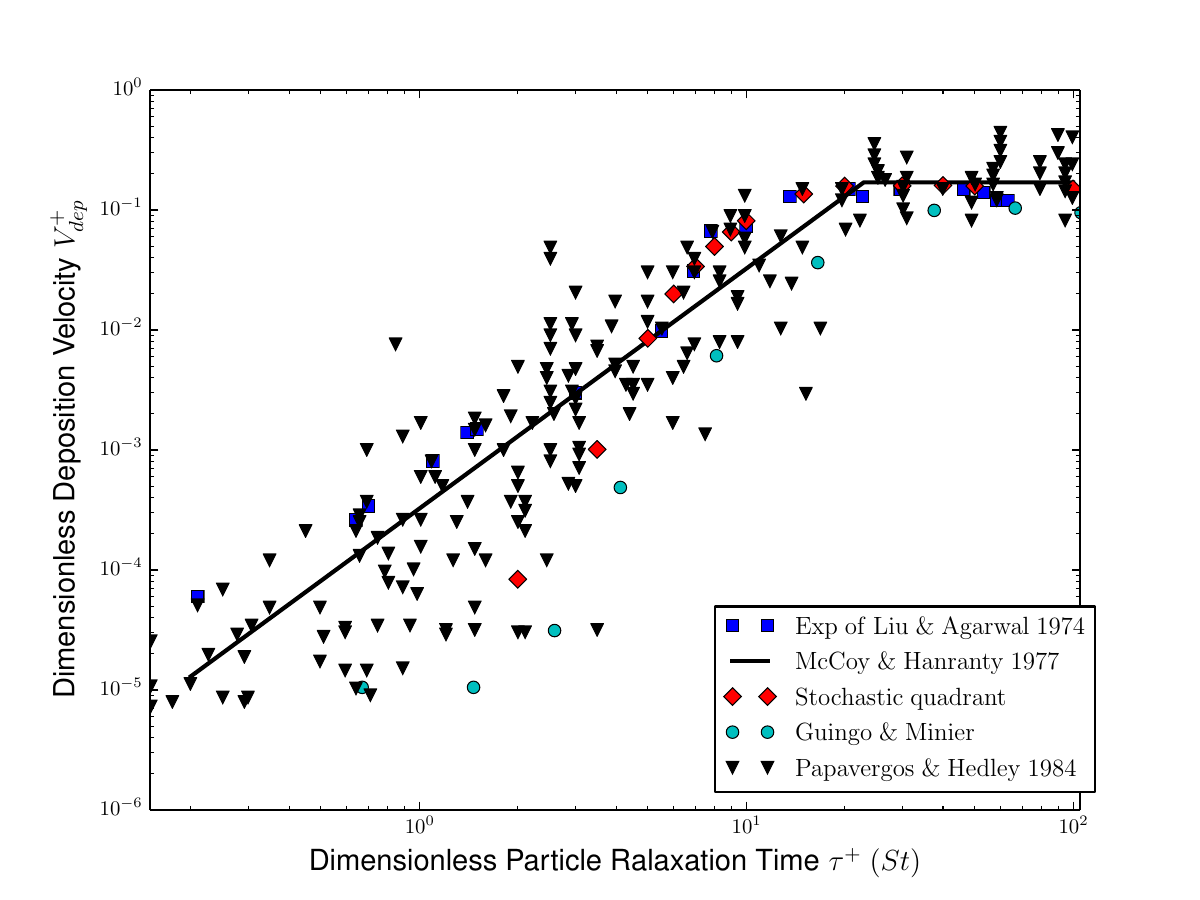} 
\caption[Comparison of dimensionless particle deposition
velocity as a function of dimensionless particle response
time with experimental measurements. ]
{Comparison of dimensionless particle deposition
velocity as a function of dimensionless particle response
time with experimental measurements.}
\label{fig:comparison_vplus2}
\end{figure}

Deposition rate data obtained in the previous study of \citeauthor{Guingo2008a}
\cite{Guingo2008a}, using a stochastic model to account for the coherent
structures, e.g. sweeps and ejections, in turbulent boundary layers,
show  good agreement with experimental data plot of \citeauthor{Papavergos1984}\cite{Papavergos1984}(who
plotted all the available measured data on deposition rates at the
time). As the present stochastic quadrant model also accounts for
such features in turbulent boundary layers, it is worth comparing our
predicted deposition rates with the same experimental data as \citeauthor{Guingo2008a}
have done. FIG.~\ref{fig:comparison_vplus2} shows the comparison.
As can be seen, the predicted deposition rates from the present stochastic
model fall well into the middle realm of experimental data of \citeauthor{Papavergos1984}
thanks to the larger scatter of the experimental data.

\subsubsection{Mean particle concentration}

\begin{figure}[htp]
\centering
\includegraphics[trim=0.00in 0.0in 0.0in
  0.0in,clip=true,scale=1.0, angle = 0, width = 1.0\textwidth]{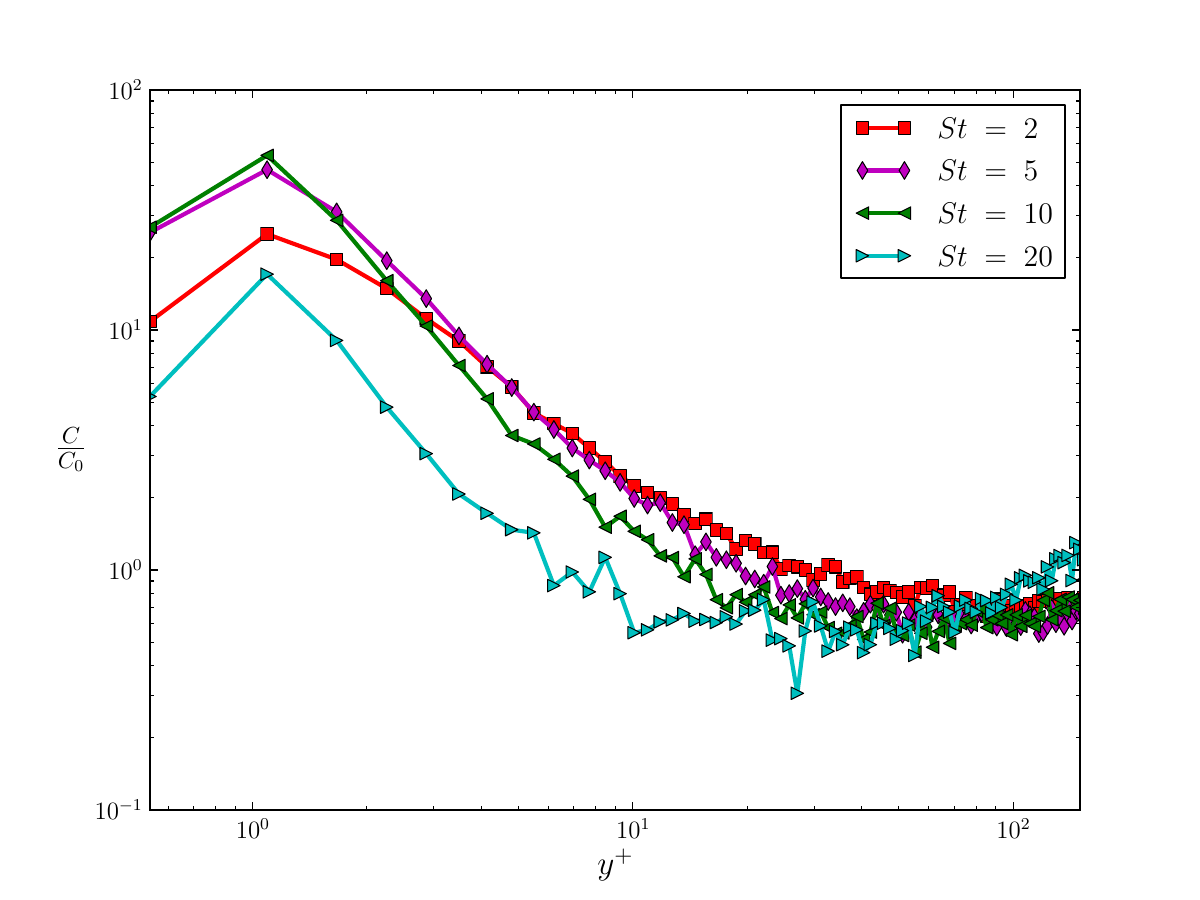} 
\caption[Particle preferential concentration profile as a
function of $y^{+}$.]
{Particle preferential concentration profile as a
function of $y^{+}$.}
\label{fig:pref_concentration}
\end{figure}

FIG.~\ref{fig:pref_concentration} shows the mean particle concentration
profile as a function $y^{+}$ across the boundary layer assuming no inter-particle collisions. There is a significant build-up in concentration for the four
classes of particles within the viscous sublayer. This build-up of particles near the wall has been observed by numerous
researchers (see~\cite{Kallio1989,Marchioli2002,Narayanan2003})
and is attributed to turbophoresis in the very near wall region (see
\cite{reeks1983transport}). The gradient in wall-normal fluid fluctuations
in boundary layer turbulence acts as a driving force and results in
a wall-ward net flux. The build-up concentration of particle with
$St=20$ is smaller than those of smaller particles with $St=2,5,10$.
This may result from the fact that these particles are too inert to follow
the relatively quiescent viscous sublayer. On the other hand, they
may move across the viscous sublayer by free flight and
deposit on the wall, which may also be responsible for the
relative reduction of build-up.

The positions of the peak concentration can be explained as a combination
of turbophoretic drift and turbulent diffusion. Very close to the
wall they enhance one another whilst further away they oppose one
another. In either case there is a constant wall-ward flux and a peak
concentration near the wall whose precise location depends upon the
divergence of the particle kinetic stresses. In the 1-D model it's
simply the gradient of particle mean square velocity. In comparing
the concentration profiles of our model with those of the Guingo and
Minier (G\&M) model \cite{Guingo2008a}, we note that the turbophoretic drift will be
significantly different because the profile of the normal kinetic
stresses in the G\&M model will be steeper. This will also influence
the local eddy diffusion coefficient. An important feature of this
balance between drift and diffusion and its dependence on particle
Stokes number, is that there will be a peaking of the particle concentration
(near the wall) and that its value and location will vary with particle
Stokes number. In particular there will be a maximum value which if
the Stokes number is defined properly in terms of the ratio of particle
relaxation time to the timescale of the local fluid motion, will occur
for $St\sim1$. Note that $St$ in this paper and elsewhere is taken
to be $\tau^{+}$, the Stokes relaxation time in wall units. The fact
that this maximum peaking with $St$ was not observed in the G\&M
concentration profiles, probably means that the G\&M had not gone far
enough in their values of $St$. This behaviour is consistent with
important features of preferential concentration which is known to
have a maximum effect for particles with Stokes numbers $\sim1$.
To our knowledge, there is no universal agreement on the 
($y^{+}$) position of the peak particle concentration. We refer to the DNS
studies by \citeauthor{Marchioli2008}\cite{Marchioli2008}, from which
the peak particle concentration seems to be at position $y^{+}<1$.
See also the results of an international collaborative benchmark test
by multiple research groups\cite{Marchioli2008} which show a similar peaking
of the concentration around $y^{+}\sim1$.

This reduction in the peak concentration and a noticeable rise in
the form of the concentration profiles near the edge of the boundary
layer for $St=20$ shown in FIG.~11 are also features of the
concentration profiles of the PDF calculations of
\citeauthor{vanDijk20124904}\cite{vanDijk20124904}, They can be explained more explicitly
in terms of a competition between turbophoresis and particle diffusion
and their individual dependence on the particle response
time $\tau_{p}$ (equivalent to $St$ when it is expressed in wall units).
The particle diffusion coefficient $\epsilon_{p}$ and the turbophoretic
drift velocity $\upsilon_{T}$ are explicitly dependent upon $\tau_{p}$
as 
\beq
\epsilon_{p}=\epsilon_{pf}+\tau_{p}\left<\upsilon_{p}^{\prime2}\right>
\qquad\upsilon_{T}=-\tau_{p}\frac{d\left<\upsilon_{p}^{\prime2}\right>}{dy}
\eeq
 where $\left<\upsilon_{p}^{\prime2}\right>$ is the particle mean square velocity
and $\epsilon_{pf}$ the particle-fluid diffusion coefficient in the
normal streamwise direction 
\cite{reeks1991kinetic,reeks2005model}. Both $\epsilon_{p}$
and $\upsilon_{T}$ are implicitly dependent upon $y$ the distance
away from the wall and also upon particle inertia through the dependence
of $\left<\upsilon_{p}^{\prime2}\right>$ and $\epsilon_{pf}$ on $\tau_{p}$ which
is essentially \emph{non-local} except for very small inertia particles.
So the equation for the concentration $\rho(y)$ (normalised by the
concentration at the edge of the boundary layer) for a constant deposition
flux depends upon the ratio of $\upsilon_{T}L/\epsilon_{p}$
\beq
d\rho/dy=\left[K-\left|\upsilon_{T}(y)\right|\rho(y)\right]L/\epsilon_{p}(y)
\eeq
where $K$ is the deposition velocity (mass transfer coefficient )
and $L$ is the effective boundary layer thickness ($\sim 100$
in wall units in our calculation). At the edge of the boundary
layer $\left|\upsilon_{T}(y)\right|/K\ll1$ , so $d\rho/dy\geq0$
and the concentration always rises as we approach the boundary layer
edge. The steepness or flatness of the curve in this region will depend
on the value of $KL/\epsilon_{p}(y)$. Both $K$ and $\epsilon_{p}$
increase with $\tau_{p}$ . So the fact that this ratio increases
with increasing $\tau_{p}$, giving a steeper curve, is because $K$
increases faster than $\epsilon_{p}/L$ as a function of $\tau_{p}$
near the boundary layer edge. \\
The peak $\rho_{max}$ will occur away from the wall for depositing
particles, and is given by 
\beq
\rho_{max}=K/\upsilon_{T}
\eeq
So although the deposition velocity $K$ increases with increasing
$\tau_{p}$ so does the turbophoretic velocity $\upsilon_{T}$. However
the ratio $K/\upsilon_{T}$ for $St=20,$ is lower than it is for
the other values of $St=2,5,10$.

\subsubsection{Mean wall-ward drift velocity and Root mean square (rms) velocity profiles}

\begin{figure}[htp]
\centering
\subfloat[]{
\includegraphics[width=0.80\textwidth]{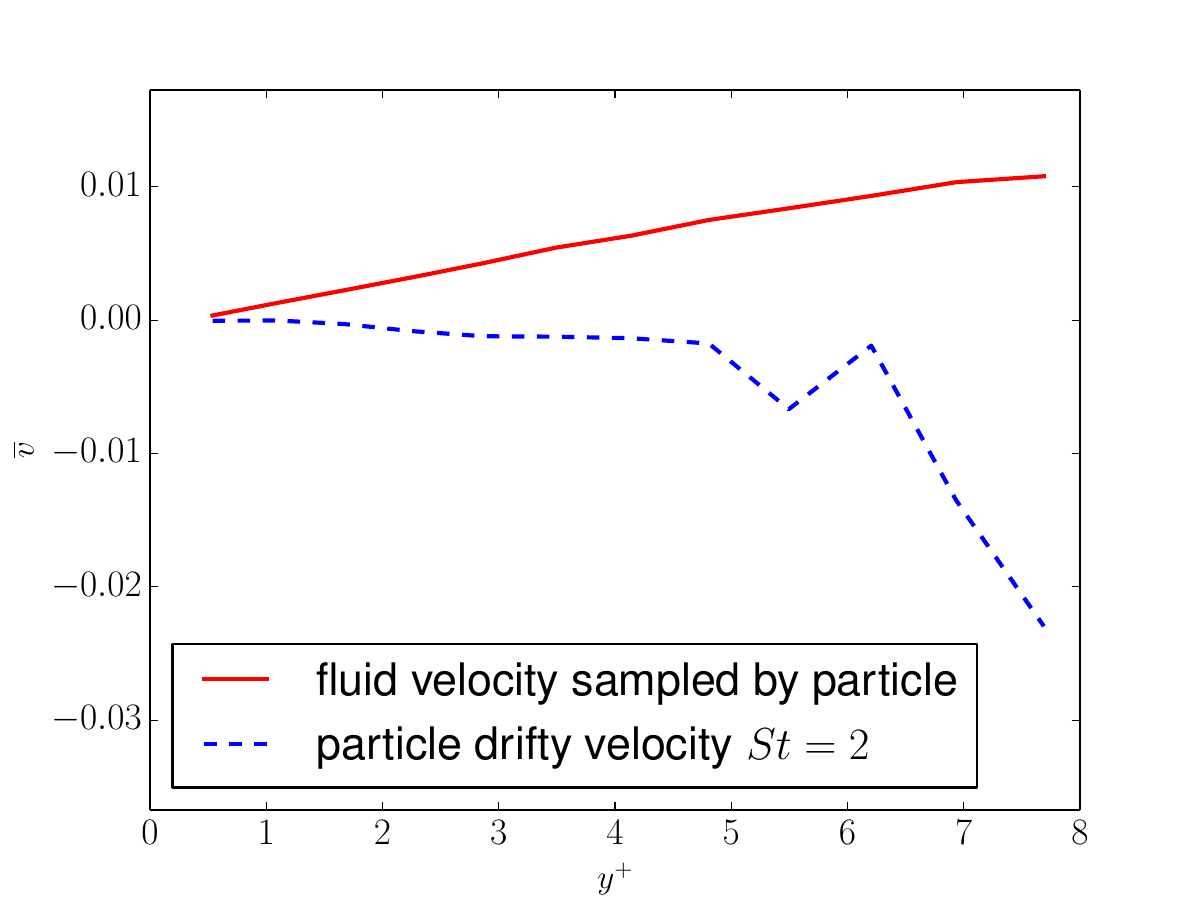}
\label{fig:subfig1_drift}}

\subfloat[]{
\includegraphics[width=0.80\textwidth]{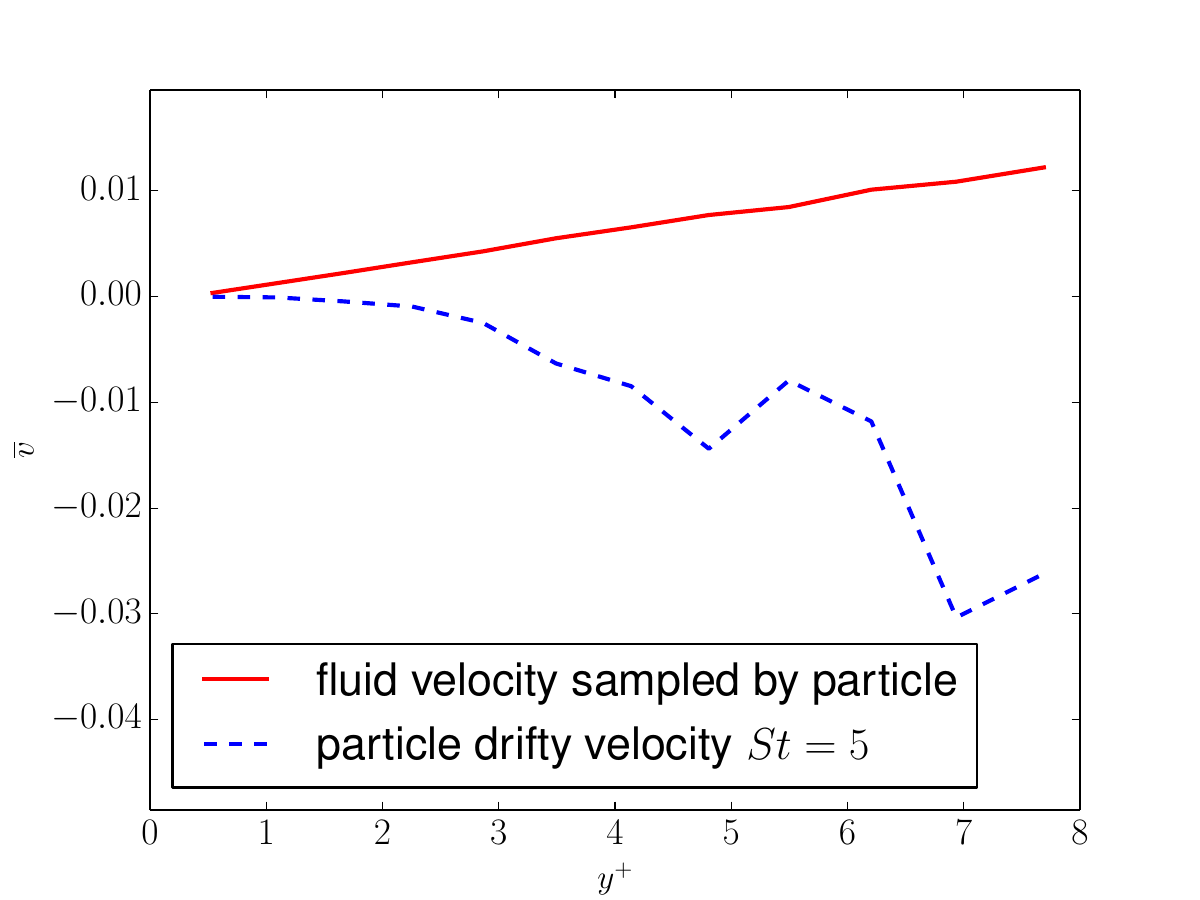}
\label{fig:subfig2_drift}}

\caption[Mean fluid wall-ward velocity sampled by particle and
particle wall-ward drift velocity $\overline{v}$,
  (a) $St = 2$, (b) $St = 5$.]
{Mean fluid wall-ward velocity sampled by particle and
particle wall-ward drift velocity $\overline{v}$, 
 (a) $St = 2$, (b) $St = 5$.}
\label{fig:glob_drift_a}
\end{figure}

\begin{figure}[htp]
\centering
\subfloat[][]{
\includegraphics[width=0.80\textwidth]{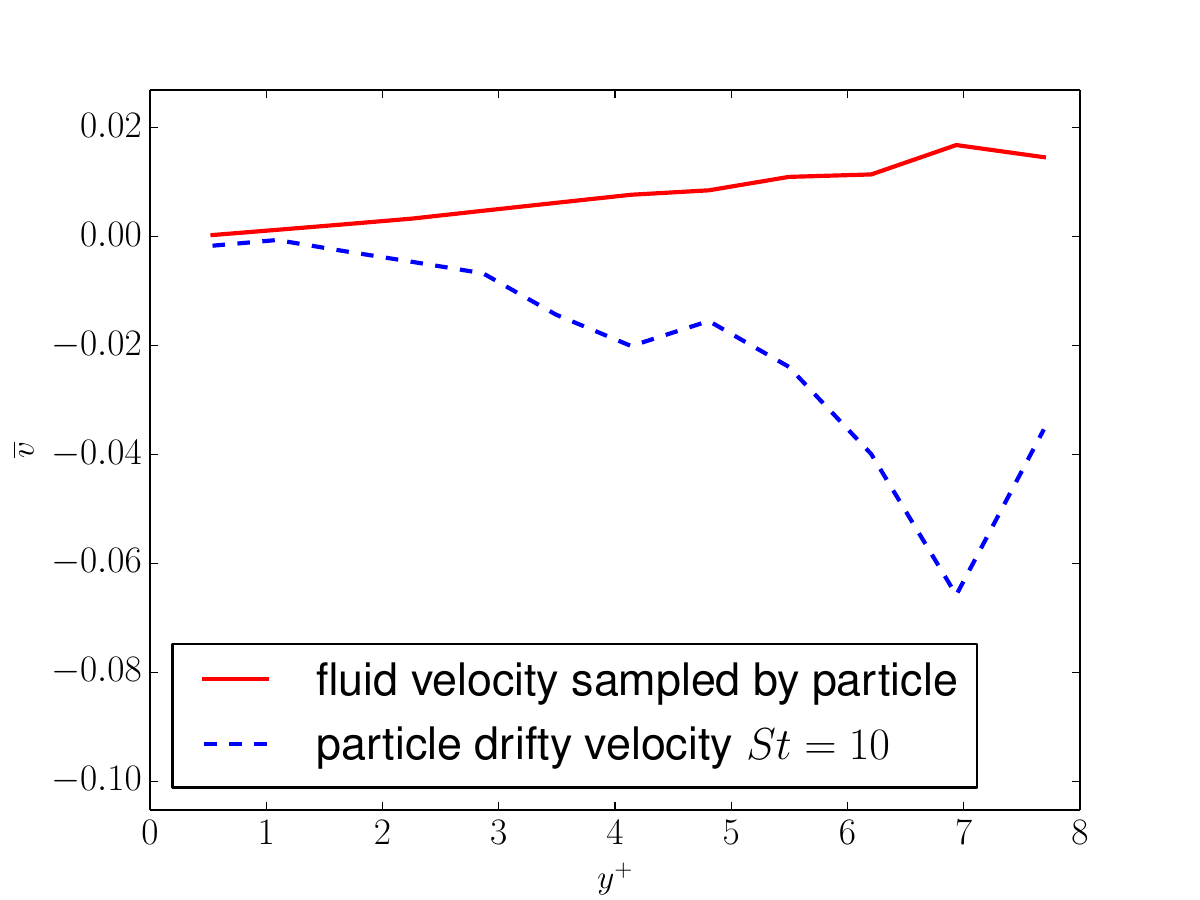}
\label{fig:subfig3_drift}}

\subfloat[][]{
\includegraphics[width=0.80\textwidth]{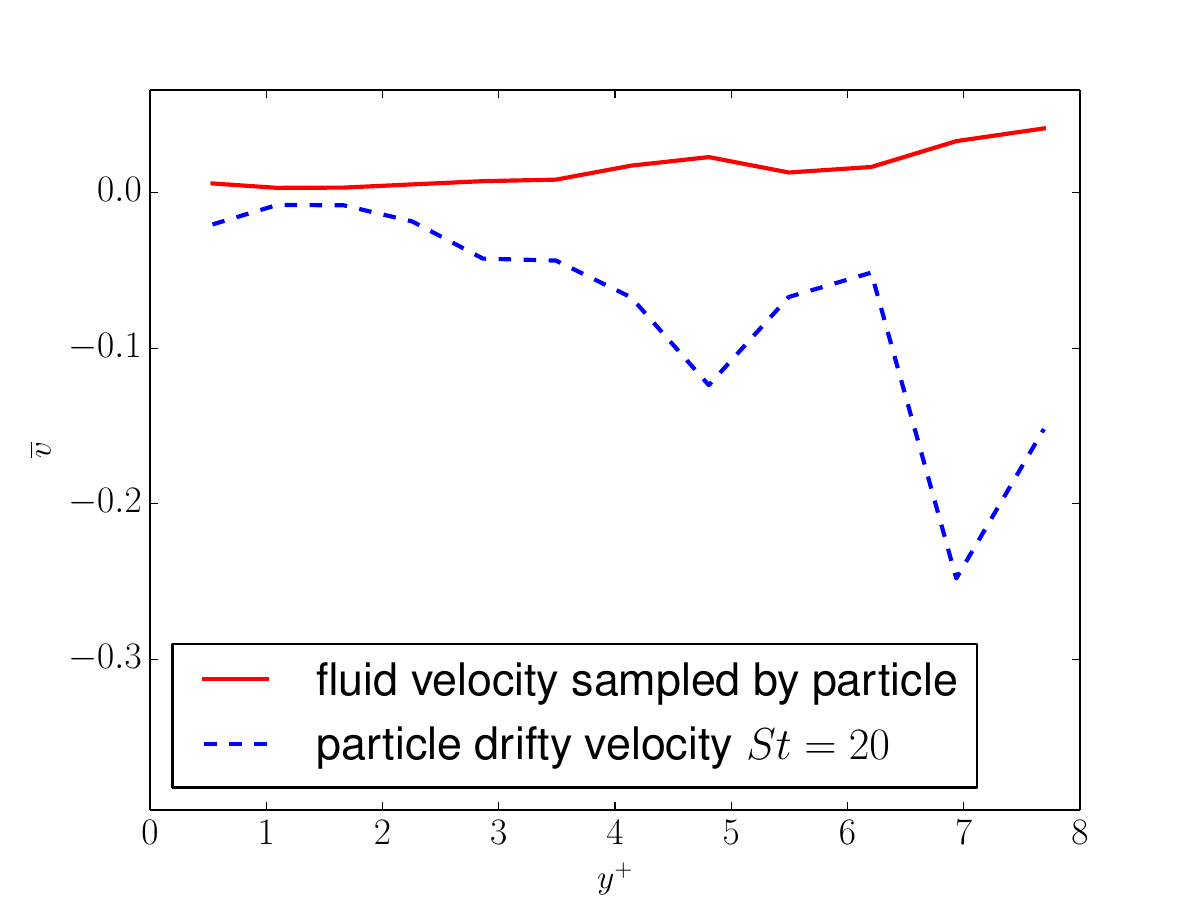}
\label{fig:subfig4_drift}}

\caption[Mean fluid wall-ward velocity sampled by particle and
particle wall-ward drift velocity $\overline{v}$,
  (a) $St = 10$, (b) $St = 20$.]
{Mean fluid wall-ward velocity sampled by particle and
particle wall-ward drift velocity $\overline{v}$, 
 (a) $St = 10$, (b) $St = 20$.}

\label{fig:glob_drift_b}
\end{figure}

FIGs.~\ref{fig:glob_drift_a} and ~\ref{fig:glob_drift_b} show the
mean wall-ward drift and sampled fluid velocity profiles in the near
wall region. We observe that the four sets of particles, $St=2,5,10,20$,
have non-zero wall-ward mean velocity (negative) values. This indicates
that the present stochastic quadrant model predicts
the phenomenon of turbophoresis. This wall-ward mean velocity of heavy
particles results primarily from the turbulence gradient of the boundary
layer turbulence as well, which is the prime mechanism responsible
for the build-up of particles. It is observed that the mean wall-ward
drift velocity of particles varies monotonically with the increase
of the particle inertia. Although the wall-normal fluid velocity has
zero mean, the sampled mean fluid velocity at the particle location
has positive values.

The existence of the wallward particle flux ties in with the particle
concentration profile which shows peaking near the wall consistent
with the existence of a turbophoretic drift towards the wall and a
diffusive flux whose direction depends upon the near wall concentration
gradient. It is clear that beyond the peak concentration, the concentration
gradient is negative with respect to $y^{+}$ indicating that the
diffusive flux is in a direction away from the wall. However the particle
flux is still directed towards the wall. It indicates that there is
a flux (independent of the concentration gradient) that is directed
towards the wall and noticeably greater than the diffusive flux. This
is of course the contribution due to turbophoresis.

The mean wall-ward fluid velocity $\overline{v}$ sampled by the particles 
is an important quantity in two-fluid modelling and much effort has
gone into its dependence on the particles' response time \cite{reeks2005model}.
Like the mean fluid velocity (sampled by a fluid element) it is generally
considered to be diffusive. The values of  $\overline{v}$  plotted
in FIGs.~\ref{fig:glob_drift_a} and ~\ref{fig:glob_drift_b} are consistent with the diffusive nature, i.e.,
when the concentration gradient is negative beyond the peak value
of the concentration, it is directed away from the wall.

\begin{figure}[htp]
\centering
\subfloat[][]{
\includegraphics[width=0.80\textwidth]{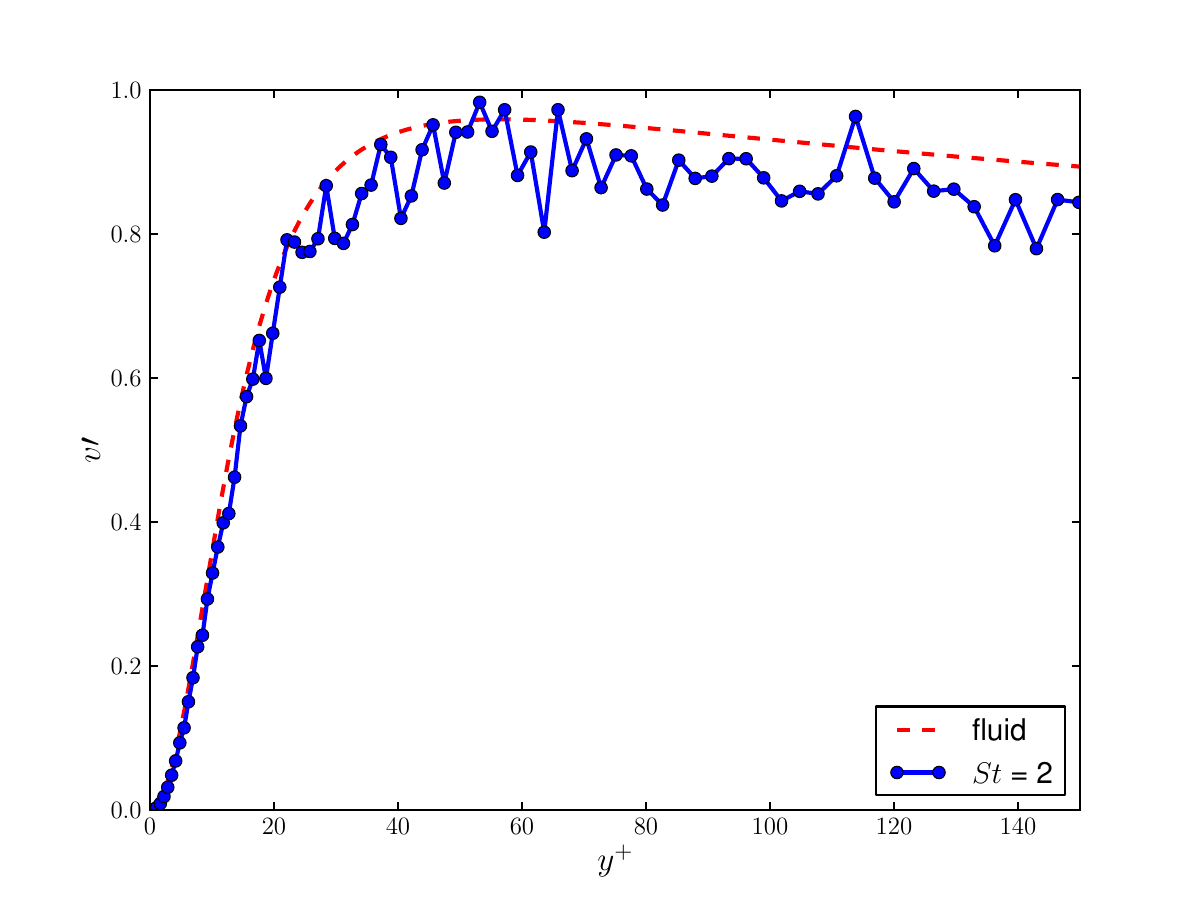}
\label{fig:subfig1_fluc_vel}}

\subfloat[][]{
\includegraphics[width=0.80\textwidth]{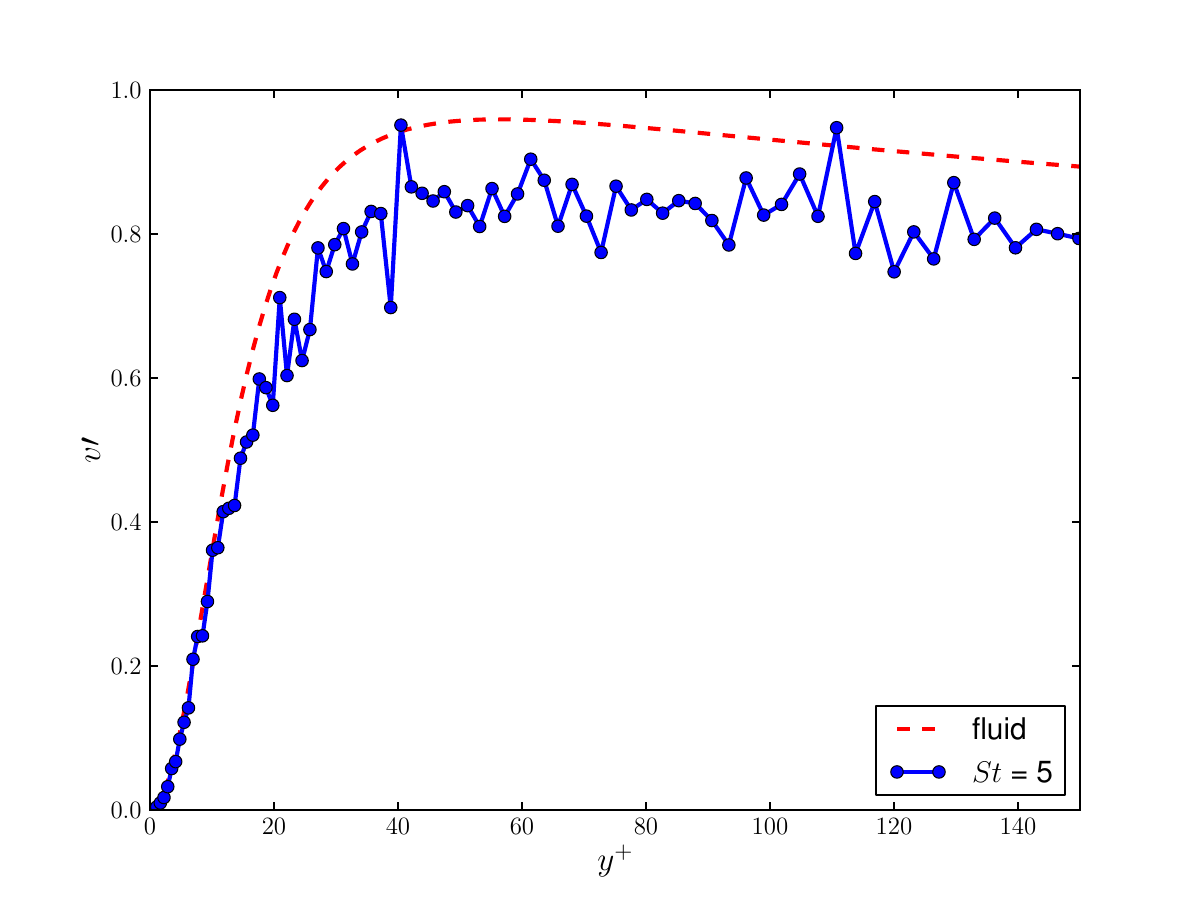}
\label{fig:subfig2_fluc_vel}}

\caption[Root mean square (r.m.s) of velocity fluctuations,
  (a) $St = 2$, (b) $St = 5$.]
{Root mean square (r.m.s) of velocity fluctuations,
  (a) $St = 2$, (b) $St = 5$.}
\label{fig:globfig_fluc_vel_a}
\end{figure}

\begin{figure}[htp]
\centering
\subfloat[][]{
\includegraphics[width=0.80\textwidth]{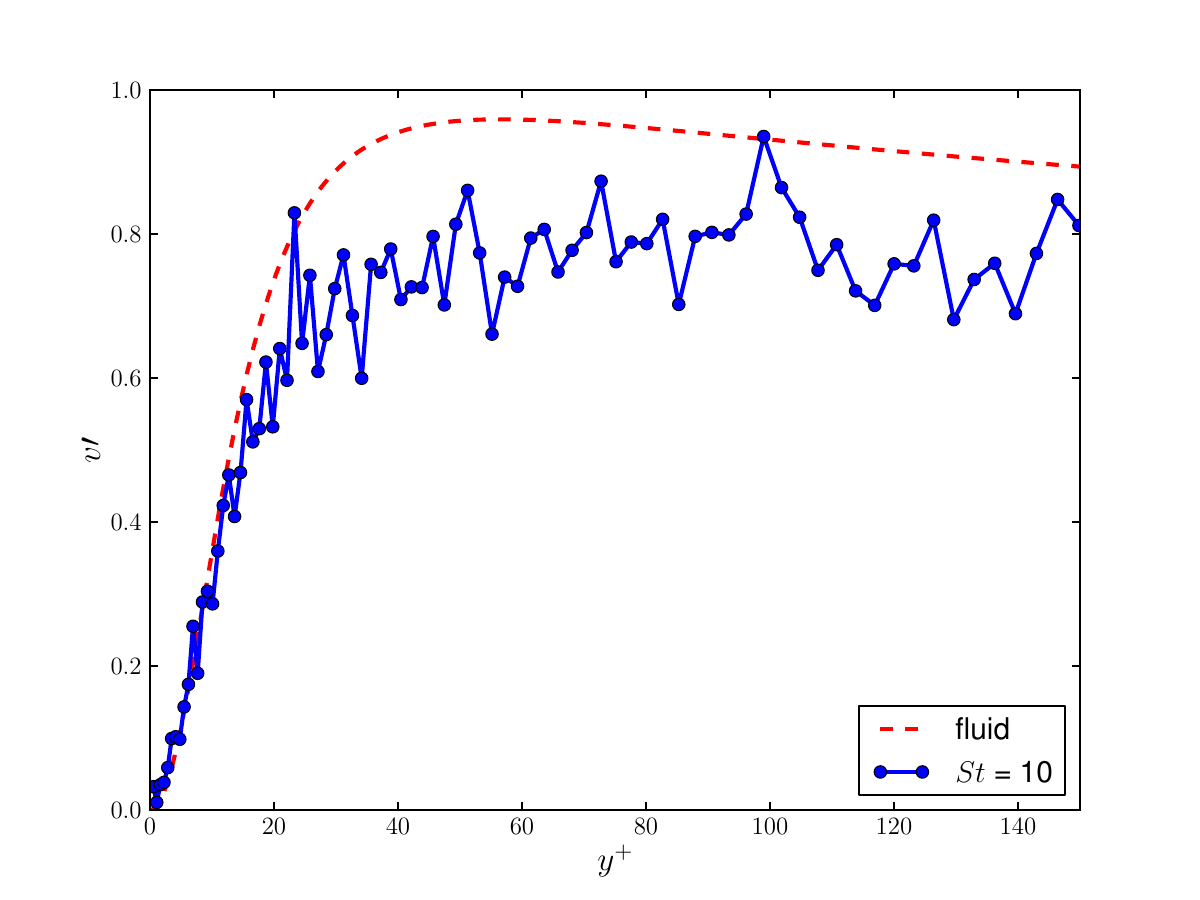}
\label{fig:subfig3_fluc_vel}}

\subfloat[][]{
\includegraphics[width=0.80\textwidth]{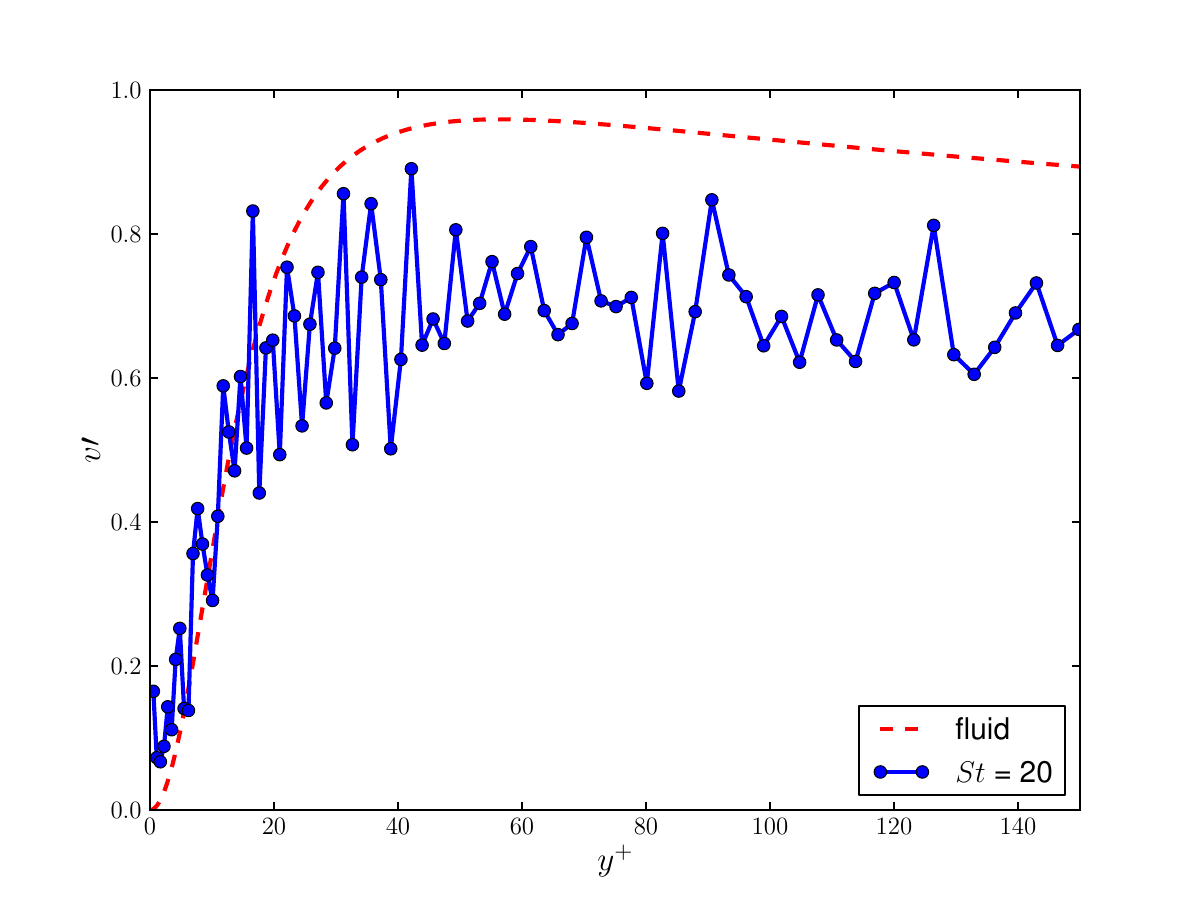}
\label{fig:subfig4_fluc_vel}}

\caption[Root mean square (r.m.s) of velocity fluctuations,
  (a) $St = 10$, (b) $St = 20$.]
{Root mean square (r.m.s) of velocity fluctuations,
  (a) $St = 10$, (b) $St = 20$.}
\label{fig:globfig_fluc_vel_b}
\end{figure}

FIGs.~\ref{fig:globfig_fluc_vel_a} and~\ref{fig:globfig_fluc_vel_b}
show the comparison of the rms of velocity fluctuations of four
sets of particles with the fluid velocity fluctuations. It is observed
that the r.m.s of particle phase is significantly different from the
fluid phase. The difference increases with increasing particle inertia.
This results from the fact that the heavier the particles, the slower
their response to the change of surrounding fluid. As far as the raggedness
displayed in the computed particle r.m.s profile is concerned, the
explanation may be that the particle phase still has not reached equilibrium
or that each sampling bin does not have a sufficient number of representative
particles.

\subsubsection{Mechanisms for particle deposition}

\begin{figure}[htp]
\centering
\subfloat[][]{
\includegraphics[width=0.80\textwidth]{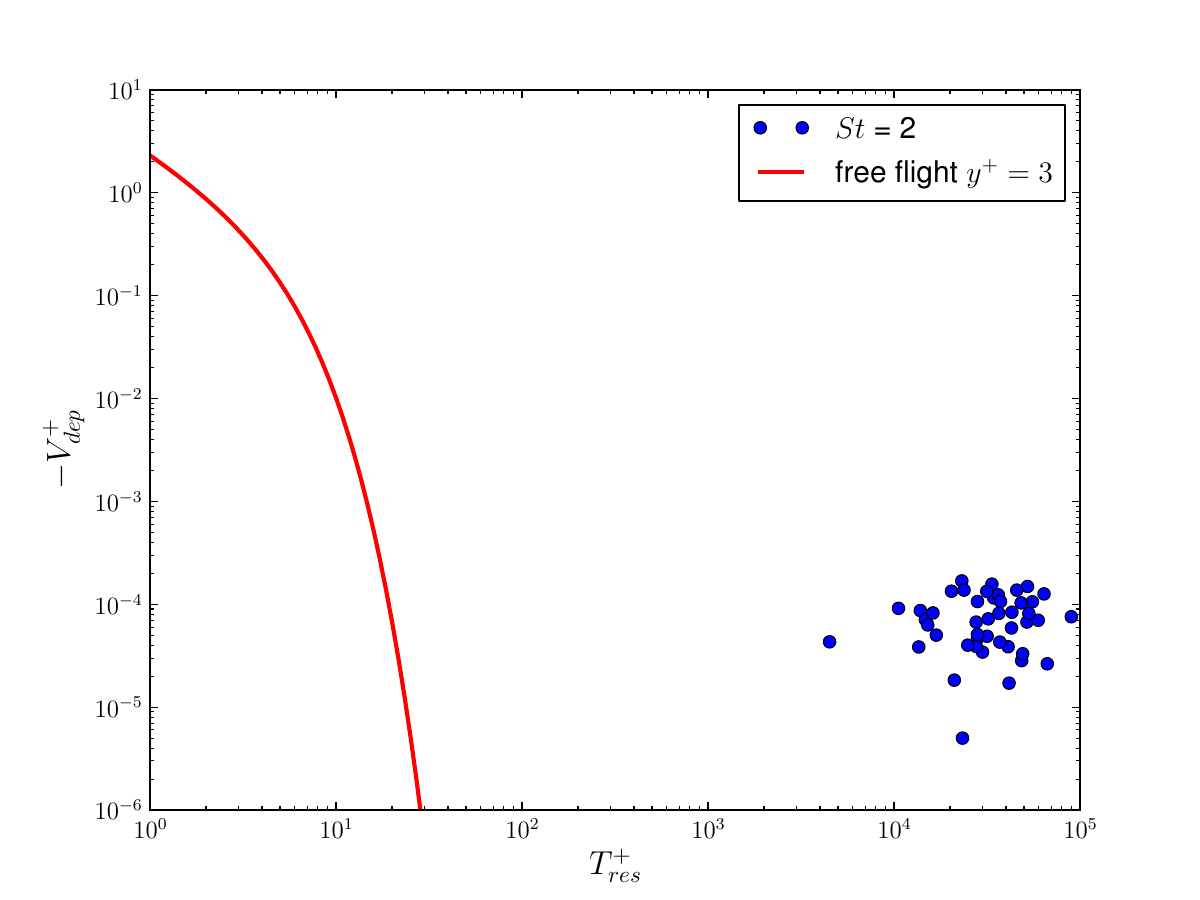}
\label{fig:subfig1_resd_T}}

\subfloat[][]{
\includegraphics[width=0.80\textwidth]{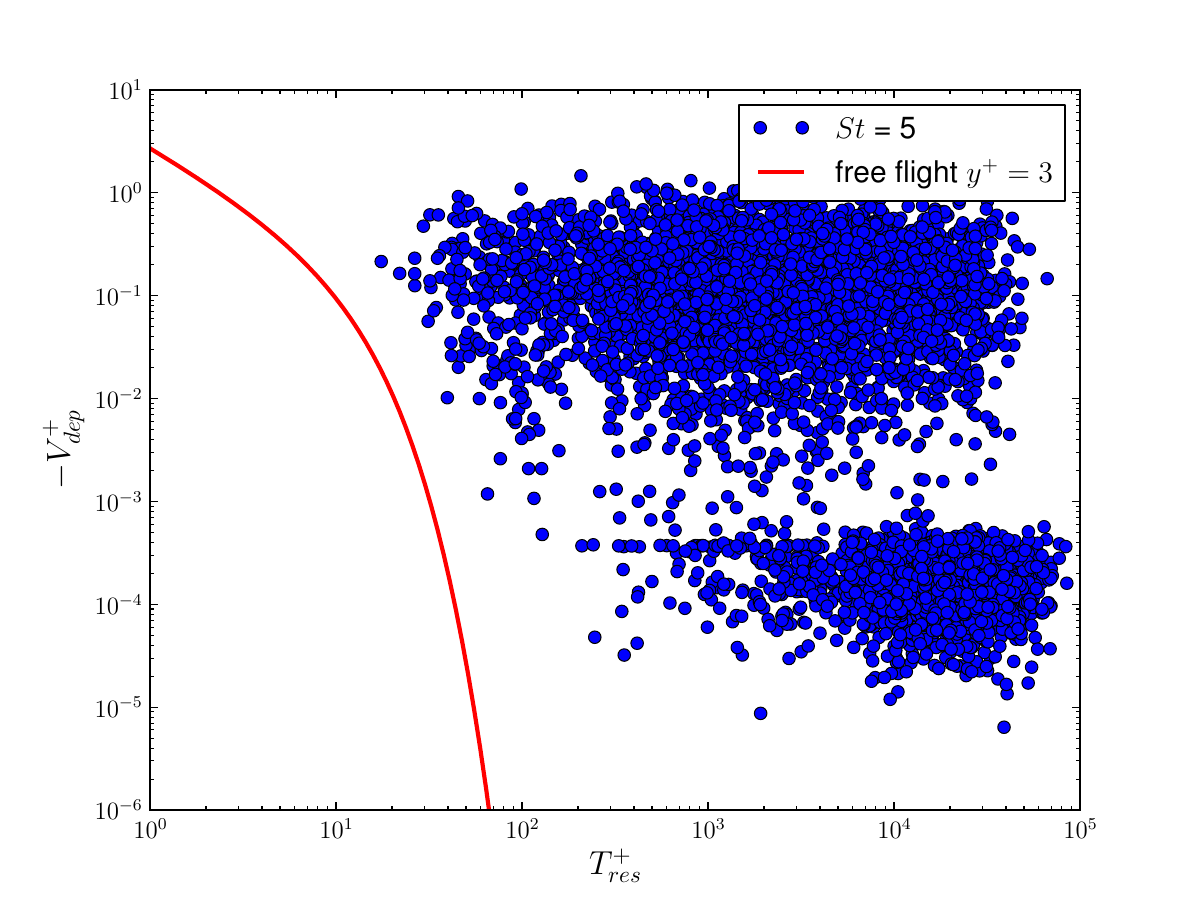}
\label{fig:subfig2_resd_T}}

\caption[Particle residence time in the region of $y^{+} <
3$ versus particle deposition velocity
  (a) $St = 2$, (b) $St = 5$.]
{Particle residence time in the region of $y^{+} <
3$ versus particle deposition velocity
  (a) $St = 2$, (b) $St = 5$.}
\label{fig:globfig_resd_T}
\end{figure}

The present stochastic model has also been used to study the mechanisms
for particle deposition. By analysing extensively the DNS data-sets
for particle transport in turbulent boundary layers, \citeauthor{brooke1994free}\cite{brooke1994free}
and \citeauthor{Narayanan2003}\cite{Narayanan2003} attributed deposition
to two different mechanisms depending on the particle inertia (Stokes
number, $St$). Relatively low inertia particles deposit by a diffusion
mechanism, whilst high inertia particles deposit as a result of free-flight.
To differentiate between the two mechanisms, the concept of particle
residence time is introduced namely the continuous time spent by a
particle within a certain wall region before depositing. Particles depositing through diffusion have relatively smaller values of deposition velocity and larger values of residence time. In contrast, particles depositing via the free-flight mechanism have the opposite values, that is relatively larger values of deposition velocity and smaller values of residence time. For the deposition
velocities and residence time, FIG.~\ref{fig:globfig_resd_T} shows
a scatter plot of wall-normal deposition velocities as a function
of particle residence time within the region of $y^{+}<3$. The red
curve is plotted according to the relation between the wall normal
deposition velocity $V_{dep}^{+}$ and the residence time $T_{res}^{+}$
provided by \citeauthor{Narayanan2003}\cite{Narayanan2003} 
\beq
V_{dep}^{+} = \frac{3 - d_p^+/2}{\tau^{+}\left[1 - \mathrm{exp}\left(\frac{T_{res}^+}{\tau^{+}}\right)\right]},
\label{eq:free_flight}
\eeq 
where $d_{p}$ is the non-dimensional
particle diameter based on the free-fight theory\cite{Friedlander1957}.
It can be observed that the deposited particles with $St=2$ do not
follow the free flight theory defined by Eq:~\ref{eq:free_flight}
as they assume relatively large near-wall residence time and
a small
deposition velocity. These particles are usually referred to as the
diffusion deposition population\cite{Narayanan2003}. For the particles
with $St=5$, there are two distinct populations of particles. The
first population assumes very long near-wall residence time and negligible
deposition velocity; however, the second population still
assumes a
relatively long near-wall residence time but relatively large deposition
velocities. The first population of particles, may be identified
as diffusion deposition particles. The near-wall residence time for
the second population is different from the DNS data calculated by
\citeauthor{Narayanan2003}\cite{Narayanan2003} as they do not follow
the free flight curve defined in Eq:~\ref{eq:free_flight}. But the
magnitude of deposition velocity of this population particles falls
into the correct range of $[10^{0},10^{-3}]$ as shown by \citeauthor{Narayanan2003}\cite{Narayanan2003}.
For all the deposited particles, the majority falls into the second
population and does not conform to the free flight particles
residence times as one might have expected. We
take that to suggest a possible alternative mechanism. As far as the over-prediction of near-wall residence time
compared to the DNS data is concerned, these particles may experience
significant repeated events both in quadrant IV (sweeps) and in quadrant
II (ejections) within the viscous sublayer, and the events in quadrant
II cause particles to be re-entrained to the outer layer or to coast along
the wall surface within the region of $y^{+}<3$ with relatively larger
velocities before being deposited. As a consequence, they assume
relatively larger velocities and larger near-wall residence time
at the same time. However it is clear that this behaviour requires
further investigation along with our proposed mechanism arising from ejection and sweeping events.

\subsubsection{A note on the influence of segregation on deposition}

This is reflected in the contribution segregation has on the net drift
velocity of particles towards the wall and it is the combination of
drift versus diffusion away from the wall that leads to a build-up
of particle concentration near the wall. The drift velocity is referred
to as turbophoresis because when it was first invoked it referred
to the migration of particles in an inhomogeneous turbulent flow from
regions of high to low regions of turbulence intensity\cite{reeks1983transport}. More precisely
the turbophoretic velocity is given by
\beq
\upsilon_{T,i}=-\tau_{p}\frac{\partial\left\langle \upsilon_{p,j}^{\prime}\upsilon_{p,i}^{\prime}\right\rangle }{\partial x_{j}}
\eeq
where $\left\langle
  \upsilon_{p,j}^{\prime}\upsilon_{p,i}^{\prime}\right\rangle $ are
the particle kinetic stresses per unit mass and $\tau_{p}$
is the particle relaxation or response
time. The formula reflects a balance between the drag and the gradients
of the particle kinetic stress at equilibrium. In the case of a fully
developed boundary layer, the drift velocity towards the wall simply
reduces to $\upsilon_{T}=-\tau_{p}d\left\langle \upsilon_{p}^{\prime2}\right\rangle /dy$.
 It has nothing to do directly with the persistence or scale of the
turbulent structures in the flow that is the direct cause of segregation.

The influence of segregation (un-mixing) has been shown to manifest
itself as an extra drift that depends upon the compressibility of
the particle velocity flow field along a particle trajectory \cite{reeks2005model}:
Reeks has referred to it as Maxey drift because it is the same expression
for the enhancement of settling under gravity due to turbulent structures
in a homogeneous turbulent flow \cite{maxey1987gravitational}. For
a flow field generated by a Langevin equation involving a white noise
driving force, the drift is zero because the flow field generated
has no structure to it (it has zero spatial correlation). The turbulent
flow field generated in this simulation does give rise to an extra
drift other than that due to turbophoresis \cite{reeks1983transport} because
it has persistence both in space and time and is spatially inhomogeneous.
However it is likely that in real boundary layer flows the combination
of vorticity and straining would lead to more pronounced segregation,
a greater enhanced drift and to greater deposition rates than predicted
by current stochastic CRW models.

\subsubsection{Probability density function (pdf) of impact velocities of particles}

\begin{figure}[htp]
\includegraphics[width=0.80\textwidth]{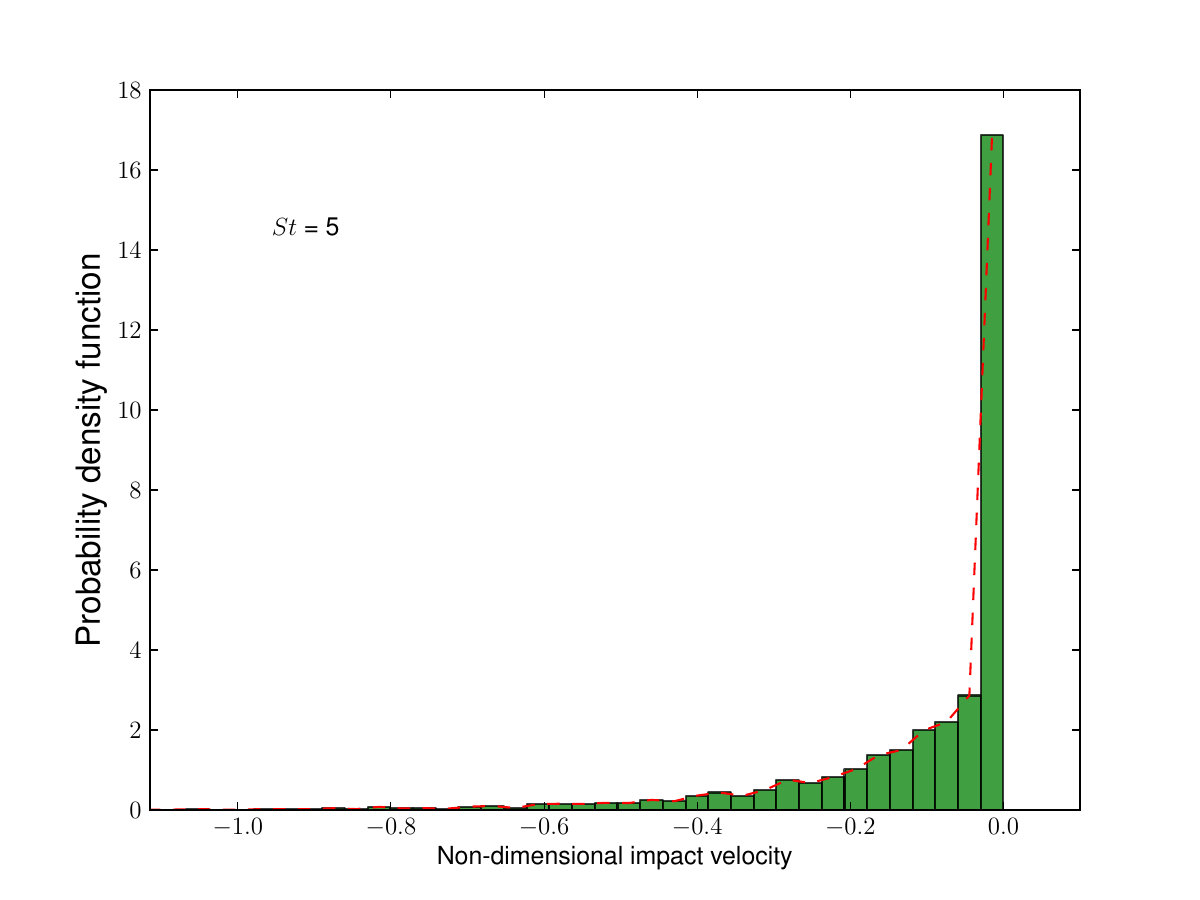}
\caption[Probability density function (pdf) of
  non-dimensional impact
  velocities of particles, $St = 5$.]
{Probability density function (pdf) of
  non-dimensional impact
  velocities of particles, $St = 5$.}
\label{fig:globfig_pdf_depV_a}
\end{figure}

\begin{figure}[htp]
\centering
\subfloat[]{
\includegraphics[width=0.80\textwidth]{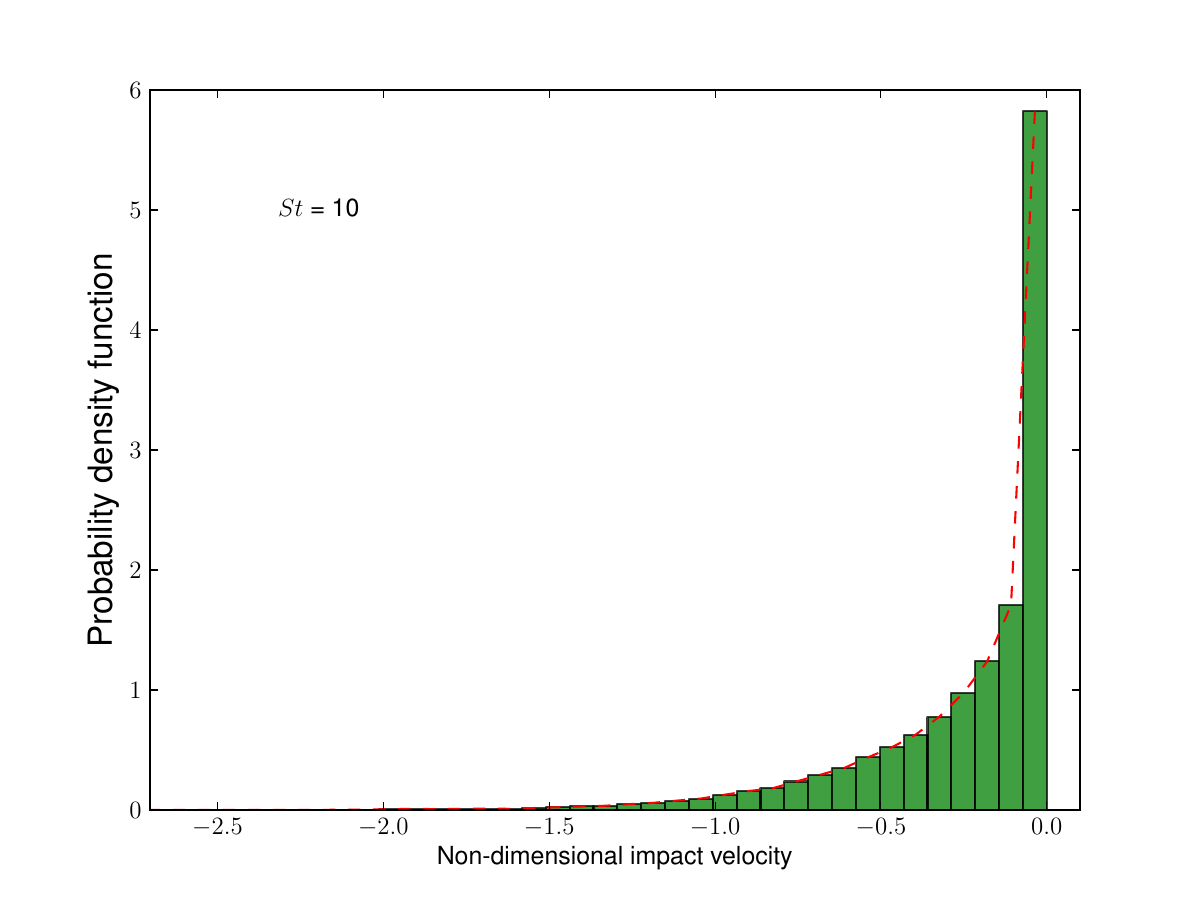}
\label{fig:subfig3}}

\subfloat[]{
\includegraphics[width=0.80\textwidth]{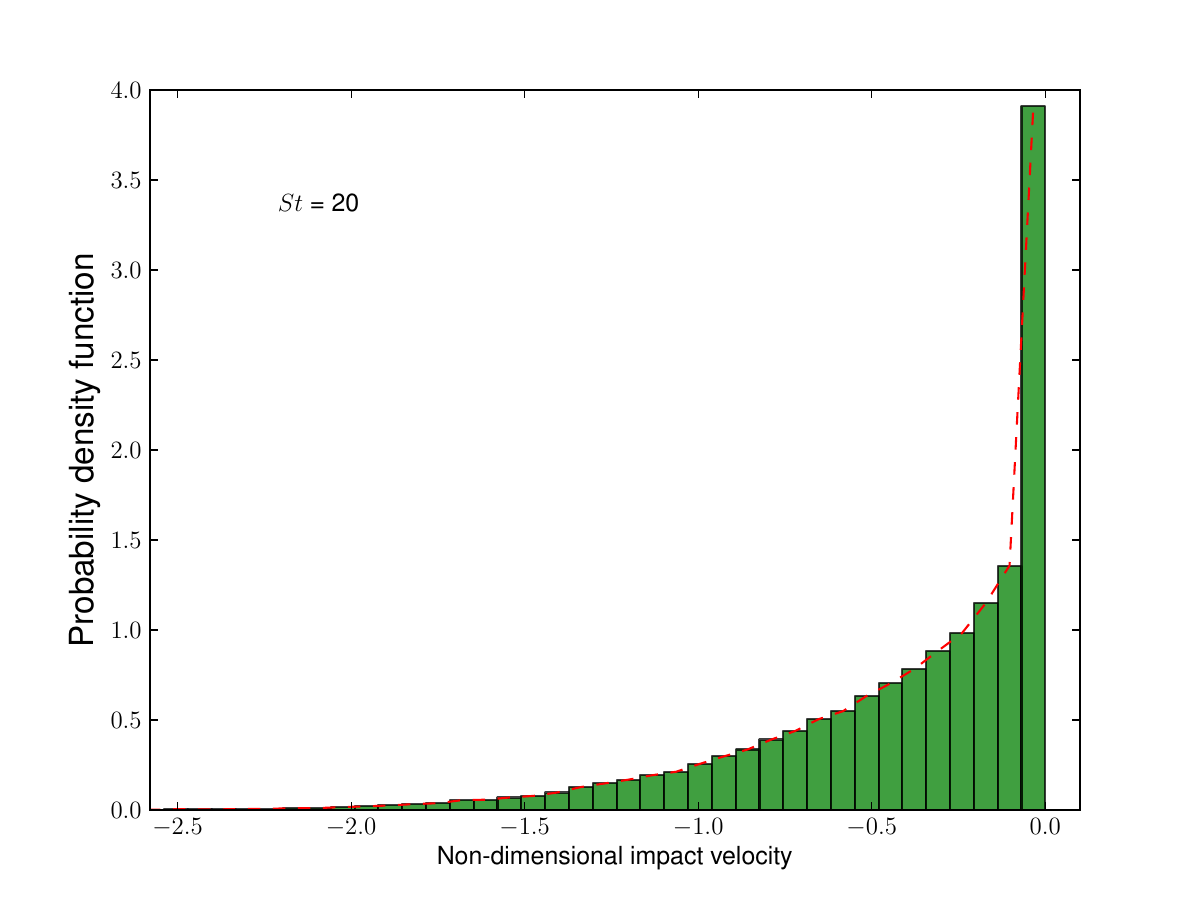}
\label{fig:subfig4}}

\caption{Probability density function (pdf) of
  non-dimensional impact
  velocities of particles,
  (a) $St = 10$, (b) $St = 20$.}
\label{fig:globfig_pdf_depV_b}
\end{figure}

FIGs.~\ref{fig:globfig_pdf_depV_a} and~\ref{fig:globfig_pdf_depV_b}
show the PDF of non-dimensional wall-normal impact velocities of depositing
particles at the wall. We see that there is a large increase
in probability in the first bin for the three sets of particles. The
particles falling in this bin may be associated with the population
of particles depositing by diffusion. There also exists a long trail
of high impact velocities, indicating some of the depositing particles
have high deposition velocities. They may be associated with free-flight
particles. The PDF of $St=20$ is much wider than those of $St=5,10$,
indicating that heavier particles are transported by free-flight across
the viscous sublayer before
deposition\cite{vanDijk20124904}.

\clearpage

\clearpage

\section{Concluding remarks}

We have described a novel stochastic quadrant model for
investigating the transport and deposition of heavy
particles in a fully developed turbulent boundary layer inspired by the quadrant analysis of
the Reynolds stress  domain by \citeauthor{Willmarth1972}\cite{Willmarth1972}.
 The detailed statistics of each quadrant are based on a quadrant analysis
of the wall-normal fluid velocity fluctuations obtained by
an LES of a fully developed channel flow. The turbulent
dispersion of heavy particles in a fully developed turbulent
boundary layer is modeled as interactions of heavy particles
with a succession of random eddies generated in four
quadrants via a homogeneous Markov chain process and was
naturally consistent with the skewness of wall-normal fluid
velocity fluctuations observed in fully turbulent developed
boundary layers. In so doing the model captures the
influence of sweeps and ejections in the near wall region of
the boundary layer on the particle deposition process.  This
model yields very good predictions of the deposition rate
for particles with $St>5$ when compared with benchmark
experimental measurements. Prediction of deposition rates at
lower values of $St$ gives a significant under-estimation
and may need further improvement. In addition, the deposition rates obtained by the
stochastic quadrant model was compared with that acquired by
solving a one-dimensional Langevin equation based on a
continuous random walk (CRW) model as well as results from
other CRW models. The discrepancy between deposition rates
for particles with $St>5$ is minor. 
When compared to  the stochastic ejection and sweep deposition model
of Guingo \& Miner \cite{Guingo2008a}, an important feature of the present model
 is its simplicity and transparency without the need for ad-hoc tuning of  model parameters.
The present data are statistically consistent with experimental analysis
of coherent structures in a turbulent boundary layer and is
much simpler to implement in RANS CFD modelling frameworks than the model
of \citeauthor{Guingo2008a}\cite{Guingo2008a}.

Most of the predicted statistics of heavy particles are consistent
with experimental measurements and DNS calculations. Build-up of particle concentration
is observed in the near wall region. This indicates that the present
stochastic model is capable of predicting turbophoresis  regarded as responsible
for this build-up. The related mean wallward drift velocity
is predicted in the viscous sublayer. Predicted  profiles of
heavy particles wall-normal r.m.s. velocity are typically lower than the
counterpart of fluid particles. Mechanisms for particle deposition
are explored by observing particle residence time versus deposition
velocity. Clearly as indicated by the residence time versus deposition velocity
shown in FIG.~\ref{fig:globfig_resd_T} in the case for $St=2$, particles reach the wall
by turbophoresis/diffusion. There are no free flight particles. The
case for FIG.~\ref{fig:subfig2_resd_T} is most interesting because although the particles
separate into two distinct populations, the regime of particles associated
with the significantly higher deposition velocities do not conform
to the free flight particles residence times as one might have expected.
We have taken that to suggest a possible alternative mechanism. However
it is clear that this behaviour requires further investigation along
with our proposed mechanism arising from ejection and sweeping events.
Studying the behaviour of higher inertial particles in the regime
where there is good agreement with experiment would be useful in establishing
when a free flight mechanism make a contribution.

The major drawbacks in the present stochastic models lie in the Lagrangian
integral time scales for the random eddies occurred in four quadrants
and in the inherent spurious drift associated with discrete random
walk models. The latter disadvantages may be corrected by introducing
an appropriate component into the particle equation of motion 
for the wall-normal fluid velocity fluctuation (see \cite{MacInnes1992}).
However, the time scales for the events in the four quadrants still call
for further investigations. 

Finally it is important to recall the concerns we expressed about
the applicability of using this approach to calculate particle deposition
in more complex flows other than in the fully developed turbulent
channel flow which has been the focus of the study we have presented
here. As with other stochastic models we imply its application in
conjunction with a RANS calculation of the underlying flow e.g $k-\epsilon$.
So in complex flows, such as over cylinders or tube-banks where the
shear stress varies rapidly around the wall surface, the model developed
in this work as with other stochastic models share the inherent wall
shear stress approximations of a RANS model employing wall functions,
which may lead to significant local errors. It is in these sorts of
flows that more detailed LES calculations of the sort carried out
for turbulent channel the LES are required.

In using LES in this way, a natural and legitimate question to ask
is ``why use a stochastic model when one can use LES directly''. The
reason is one of computational efficiency or perhaps more precisely
a matter of reliability and statistical accuracy. There is a lot of
statistical information contained in an LES calculation which
is not wholly relevant when we use it to calculate particle deposition.
So the construction and application of a simple stochastic model of
the sort described and studied in this paper has been to extract
the most relevant and most appropriate information about near wall
turbulent flow that is important in accurately predicting deposition
and to incorporate this into a model in the most computationally efficient
way. Such a model can then be used to calculate particle deposition
for a whole range of particles sizes, flows and particle wall boundary
conditions (from perfectly absorbing to partially absorbing).

\begin{acknowledgments}
  The financial support and advice of British Energy (part
  of EDF Energy, Barnwood, Gloucester, U.K.) is gratefully
  acknowledged. However the views expressed in this paper
  are those of the authors and do not necessarily represent
  the views of the sponsors.  The authors would also like to
  thank Dr. David C. Swailes for the insightful discussions
  on the development of the stochastic quadrant
  model. Finally, we wish to acknowledge the
  anonymous reviewers for useful feedback on the revision of
  this paper.

\end{acknowledgments}

\clearpage
\bibliography{Deposition_PoF}

\end{document}